\documentclass[12pt]{iopart}

\usepackage{iopams} 
\usepackage{color}
\usepackage{graphicx} 
\usepackage{sidecap} % enables caption *next* to figure, if we need it

\newcommand{\gsim}{\lower.7ex\hbox{$\;\stackrel{\textstyle>}{\sim}\;$}}
\def\hmpcinv{\,h\,{\rm Mpc^{-1}}}

\newcommand{\Nside}{N_{\rm side}}
\newcommand{\ellmax}{\ell_{\rm max}}
\newcommand{\Ylm}{Y_{\ell m}}

\renewcommand*{\vec}[1]{\bi{#1}}
\newcommand*{\unitvec}[1]{\hat\vec{#1}}

\begin{document}

\title[CMB anomalies after Planck]{CMB anomalies after Planck}

\author{ Dominik J Schwarz$^1$, Craig J Copi$^2$, Dragan Huterer$^3$, and Glenn D Starkman$^2$}
%\author{Craig J Copi$^1$, Dragan Huterer$^2$, Dominik J Schwarz$^3$, and Glenn D Starkman$^1$}

\address{$^1$ Fakult\"at f\"ur Physik, Universit\"at Bielefeld,
 Postfach 100131, 33501 Bielefeld, Germany}
\address{$^2$ CERCA/Department of Physics/ISO, Case Western Reserve University,
 Cleveland, OH 44106-7079, USA}
\address{$^3$ Department of Physics, University of Michigan, 
 450 Church St, Ann Arbor, MI 48109-1040, USA}

\ead{\mailto{dschwarz@physik.uni-bielefeld.de}, 
\mailto{cjc5@cwru.edu}, \mailto{huterer@umich.edu}, 
\mailto{glenn.starkman@case.edu}}

\begin{abstract}
Several unexpected features have been observed in the microwave sky at large
angular scales, both by WMAP an by Planck.  Among those features is a lack of
both variance and correlation on the largest angular scales, alignment of the
lowest multipole moments with one another and with the motion and geometry of
the Solar System, a hemispherical power asymmetry or dipolar power modulation,
a preference for odd parity modes and an unexpectedly large cold spot in the
Southern hemisphere.  The individual p-values of the significance of these
features are in the per mille to per cent level, when compared to the
expectations of the best-fit inflationary $\Lambda$CDM model.  Some pairs of
those features are demonstrably uncorrelated, increasing their combined
statistical significance and indicating a significant detection of CMB
features at angular scales larger than a few degrees on top of the standard
model.  Despite numerous detailed investigations, we still lack a clear
understanding of these large-scale features, which seem to imply a violation
of statistical isotropy and scale invariance of inflationary perturbations.  In
this contribution we present a critical analysis of our current understanding
and discuss several ideas of how to make further progress.
\end{abstract}


\submitto{\CQG}

\maketitle


\section{Introduction}

Among the purposes of this contribution is to summarize the evidence for unexpected features of the microwave sky 
at large angular scales, as revealed by the observation of temperature anisotropies by the space missions
Cosmic Background Explorer (COBE), Wilkinson Microwave Anisotropy Probe (WMAP) and Planck. 
Before doing so, let us put those discoveries into context with the study of other aspects of the cosmic 
microwave background (CMB) radiation.   

Half a century ago, the discovery of the CMB revealed 
that most of the photons in the Universe 
belong to a highly isotropic thermal radiation at a temperature of $\sim3$K~\cite{CMBdiscovery}. 
Deviations from this isotropy were first found 
in the form of a temperature dipole at the level of $\sim3$mK \cite{CMBdipole1,CMBdipole2}. 
This dipole has been interpreted as the effect of Doppler shift and aberration 
due to the proper motion of the Solar System \cite{CMBProperMotion} with respect to a
cosmological rest frame.

The observation of an isotropic CMB, together with the proper-motion
hypothesis, provides strong support for the cosmological principle.  This
states that the Universe is statistically isotropic and homogeneous, and
restricts our attention to the Friedmann-Lema\^itre class of cosmological
models.  The cosmological principle itself is a logical consequence of the
observed isotropy and the Copernican principle, the statement that we are
typical observers and thus observers in other galaxies should also see a
nearly isotropic CMB.

The proper-motion hypothesis is supported by the COBE discovery of higher multipole moments 
\cite{COBE}. These higher moments turned out to be two orders of magnitude below the dipole signal,
at a rms temperature fluctuation of $\sim 20 \mu$K at COBE resolution. 
However, a direct test of the proper-motion hypothesis had to wait 
until Planck was able to resolve the Doppler shift and aberration of hot and cold spots 
at the smallest angular scales \cite{PlanckProperMotion, Planck2015Isotropy}. 
It is important to note here that the observed dipole could also receive contributions 
from effects other than the Solar System's proper motion. 
These could be as large as 40 per cent without contradicting
the Planck measurement at the highest multipole moments. 
Observations at non-CMB frequencies, e.g. in the radio or infra-red, 
hint at significant structure dipoles or bulk flows, 
but are still inconclusive 
\cite{BlakeWall,Singal2011,GibelyouHuterer2012,RubartSchwarz2013,Kothari2013,
Singal2014,Yoonetal2014,TiwariJain2015,TiwariNusser2015}. 
Here we dwell on this aspect as the CMB dipole is one of the 
most important calibrators in modern cosmology. 
It defines what we call the CMB frame and many 
cosmological observations and tests refer to it. 

The existence of structures like galaxies, voids and clusters imply that the
CMB cannot be perfectly isotropic.  The COBE discovery \cite{COBE} revealed
the long-expected temperature anisotropies and confirmed that they are
consistent with an almost scale-invariant power spectrum of temperature
fluctuations.
%This was consistent with the idea of a quantum origin of cosmic fluctuations. \dominik{Should the quantum 
%fluctuations stay here? Does they require a citation here? This point is explained in more detail below.}
Scale invariance of the temperature anisotropies means that the
%\newline \dominik{band?}  \glenn{why is this band?} \dominik{actually
%$C_\ell$ is the angular power spectrum, $D_\ell$ is the band power in a
%logarithmic interval in $\ell$, thats the same as $P(k)$ being the power
%spectrum and $k^3/2 \pi^2 P(k)$ being the band power for 3 d.}
band power spectrum $D_\ell \equiv \ell (\ell + 1) C_\ell/2\pi$ 
is a constant for small multipole number $\ell$. 
Here $C_\ell$ denotes the expected variance 
in the amplitude of any spherical harmonic component of the temperature fluctuations
with total angular-momentum $\ell$.\footnote{This analogy from 
quantum physics is useful to describe the spherical harmonic analysis of temperature 
fluctuations in terms of well-known physical concepts.}

\begin{figure}
\hfill \includegraphics[width=0.8\linewidth]{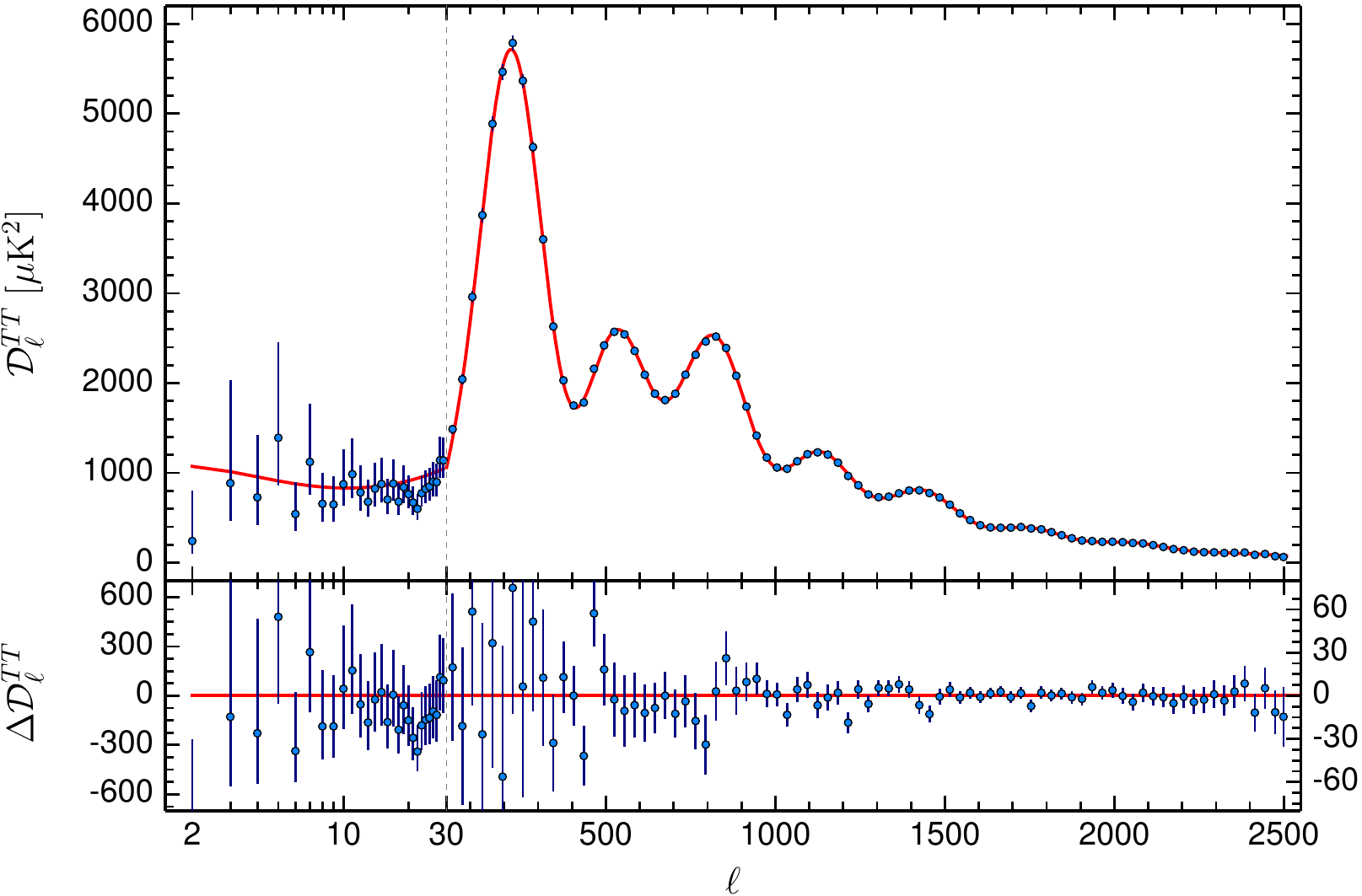}
\caption{Angular band power (top) and residual angular band power (bottom) of
  the cosmic microwave temperature anisotropies as presented in the Planck
  2015 release \cite{Planck2015Overview}. The error bars show the sum
    of measurement error and cosmic variance, the latter being the dominant source
  of uncertainty at large angular scales.}
\label{Planck2015_TT}
\end{figure}

During the last two decades, ground-based, balloon-borne and satellite CMB
experiments led to an improved understanding of those temperature
anisotropies.  The WMAP and Planck space missions played a special role, 
obtaining full-sky measurements that
enabled us to investigate a large range of angular scales, from the dipole
$\ell =1$ to $\ell \sim 2500$, more than three decades in $\ell$.  The band
power spectrum as published by Planck is shown in Fig.~\ref{Planck2015_TT}.

These temperature fluctuations are believed to have been generated 
from quantum fluctuations in the very early Universe
\cite{MukhanovChibisov1981} by a (nearly) scale-invariant mechanism. 
The most prominent context is cosmological inflation \cite{Starobinsky1980,Guth1981}. 
If inflation lasts long enough, 
the spatial geometry of the Universe is generally predicted to be indistinguishable from Euclidean,
and the topology of the observable Universe is expected to be trivial (simply connected).
Even more importantly, 
inflation predicts that the CMB temperature fluctuations should be:
(i) statistically isotropic, 
(ii) Gaussian, 
and (iii) almost scale invariant.
It also predicts:
(iv) phase coherence of the fluctuations;
(v) for the simplest models, a dominance of the so-called adiabatic mode 
(strictly speaking it is not only adiabatic but also isentropic); 
and (vi) the non-existence of rotational modes at large scales. 
Finally, depending on the energy scale of cosmological inflation, 
there might be (vii) a detectable stochastic background of gravitational waves \cite{Starobinsky1979} 
that also obeys properties (i) to (iii). 

In the process of extracting cosmological parameters from the CMB and other observations, 
properties (i) to (vi) are assumed to hold true and a stochastic gravitational wave background is 
neglected. This leads to the minimal inflationary $\Lambda$ cold dark matter ($\Lambda$CDM) model 
\cite{Planck2015Parameters}. 

Analysis of the CMB allows us not only to fit all free parameters of this
model, but also to test its underlying assumptions. However, the more
fundamental the assumption, the harder it appears to test. The existence of
the peaks and dips shown in Fig.~\ref{Planck2015_TT} are due to the phase
coherence, property (iv).  The almost-scale-invariance (iii) is visible in the
smallness of the deviations from the best-fit model, although a
model-independent reconstruction of the primordial power spectrum leaves room
for deviations at the largest observed scales \cite{Planck2015Inflation}.
More detailed analysis also reveals that there is a strong upper limit of at
most $4\%$ of non-adiabatic modes (v) \cite{Planck2015Inflation}, while
rotational modes would have produced a large B-polarization signal that is not
observed.  The predicted flatness and the expected trivial topology are
consistent with all observations
\cite{Planck2015GeometryTopology,Vaudrevange:2012da}.

It thus remains to test Gaussianity and statistical isotropy. 
A lot of effort has been put into searches for non-Gaussianity 
and they are described in great detail elsewhere. 
The brief summary is that there is no evidence for it so far \cite{Planck2015Gaussianity}. 
In the following we focus our attention on statistical isotropy,
and touch on the issue of scale invariance.    

\begin{figure}
\hfill \includegraphics[width=0.8\linewidth]{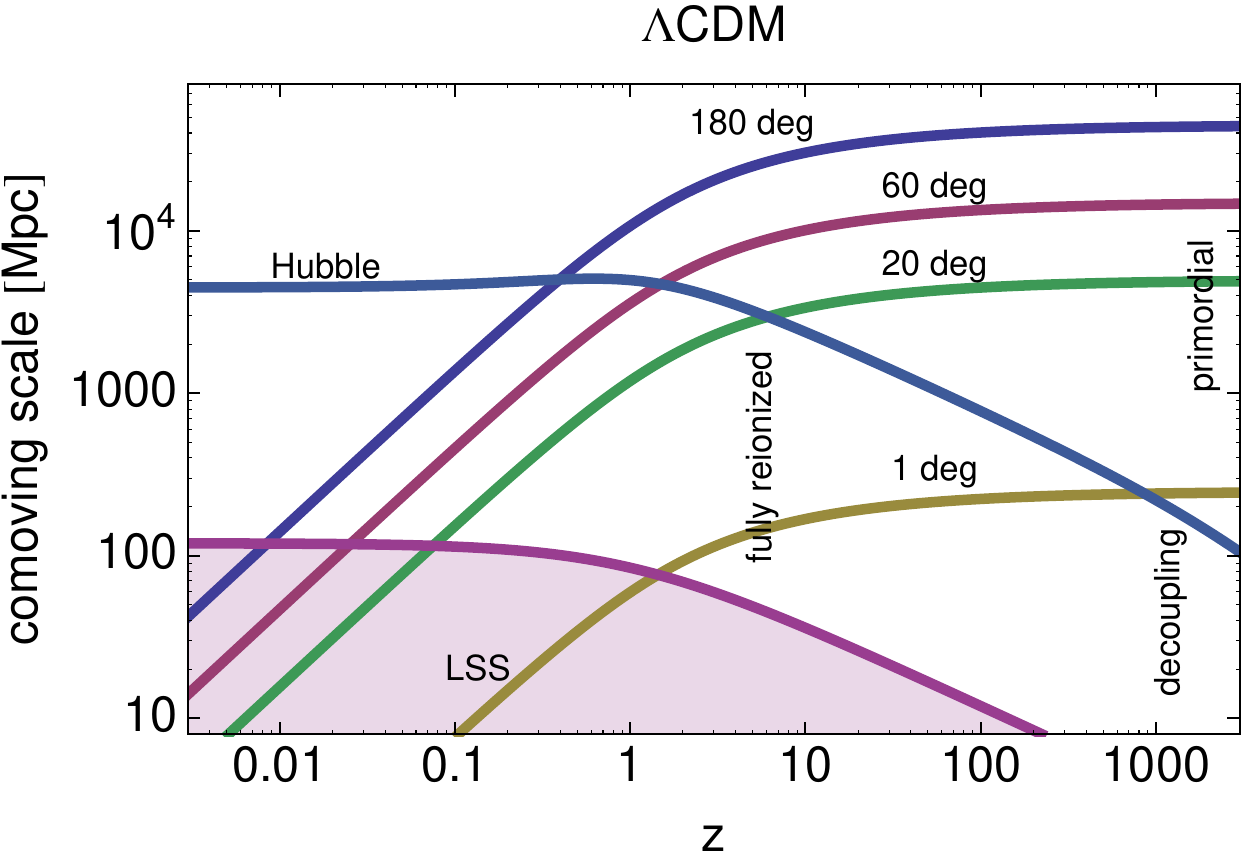}
\caption{The comoving length of an arc on the sky with opening angle of $180,
  60, 20$ and $1$ degrees is compared to the comoving Hubble distance. Angular
  scales larger than $60 (20, 1)$ degrees can only be affected by primordial
  physics or physics at redshift $z < 1 (10, 1000)$, corresponding to the
  present (the reionized, the transparent) universe. The shaded region
  indicates scales and redshifts on which structure formation is expected to
  generate density contrasts of order $0.1$ or larger. \label{AngularScales}}
\end{figure} 

All mentioned predictions should hold at all observable scales. However, testing these primordial properties of the 
Universe directly is complicated by physics related to the evolution of the Universe after the end of cosmological 
inflation. In order to understand which phenomena can be most cleanly probed at which scales it is instructive to 
look at the comoving size corresponding to a particular angular scale as a function of redshift, see figure 
\ref{AngularScales}. 

At the time of the formation of the first atoms, 
scales that today subtend more than about a degree (and that therefore affect $\ell \lesssim 180$)
were not much affected by details of photon decoupling. 
Thereafter, the Universe was filled with a mix of H and He gas, 
until it was reionized at a redshift of about 10. 
Angular scales larger than about 20 degrees (or $\ell \lesssim 10$) 
are also not much affected by the details of reionization. 
Finally, angular scales larger than $\sim 60^\circ$  (or $\ell \lesssim 3$) 
enter the Hubble scale at a redshift of one and thus are either of primordial or local origin. 
Here by local we mean from within our Hubble patch of the Universe. 
Thus it is a good idea to start with a test of statistical isotropy at the
largest angular scales, as whatever we find must be either primordial or a local effect due to either foreground or 
local cosmic structure. 

In this contribution we intend to give a summary of the evidence for the
existence of features of the microwave sky that apparently violate statistical
isotropy on the largest angular scales (section 2).  Since this seems to
happen only at the largest angular scales, it also amounts to a violation of
scale invariance.  We also discuss several ideas that have been put forward to
explain those features, though we do not intend to give an exhaustive review.
Apart from the suggestion that all of them are statistical flukes (the
probability for which to happen is tiny, unless compensated for by huge
look-elsewhere penalties) these ideas can be classified into foreground
effects (section 3) and cosmological effects (section 4).  In section 5 we
highlight several possible tests of those ideas.  The study of polarization at
large angular scales and more detailed all-sky study of non-CMB wavebands
seems to be particularly promising.

\section{A summary of the evidence} 

Some of the unexpected features in the CMB temperature anisotropies have been identified in angular space, 
$T(\unitvec{e})$, where $\unitvec{e}$ is a unit vector describing a position on the sky, 
and some in harmonic space, 
\begin{equation}
a_{\ell m} = \int Y^*_{\ell m} (\unitvec{e}) T(\unitvec{e}) {\rm d}\unitvec{e},
\end{equation} 
where $Y_{\ell m}(\unitvec{e})$ denote spherical harmonic functions. 
%If we had perfect observations on a full sky, both methods would be completely equivalent. 
Mathematically, all of the information in a full-sky map, $T(\unitvec{e})$,
is contained in the $a_{\ell m}$. 
However, as one often finds in transforming data, 
the different representations  can reveal complementary features.
Furthermore, many of the usual relationships between angular-space and harmonic quantities are 
complicated by the presence of the galactic foreground, 
which  forces us to mask, weight and clean the observed maps. 
%Many of these procedures are non-linear and spoil the 
%one-to-one equivalence of angular and harmonic space. 
Thus it is advisable and fruitful to study both sides of the spherical-harmonic coin. 

For a statistically isotropic sky, the one-point expectation values are 
$T_0 = \langle T(\unitvec{e})\rangle$ and 
$\langle a_{\ell m} \rangle = \delta_{\ell 0}\delta_{m 0} \sqrt{4 \pi} T_0$, both 
quantities with an arbitrarily large cosmic variance, thus $T_0$ cannot be predicted. It is a free parameter
of the $\Lambda$CDM model and must be measured, which was first done by Penzias  and Wilson 
\cite{CMBdiscovery}
and most accurately to date  by COBE \cite{FixsenMather2002,Fixsen2009}.  
We also employ the usual hypothesis of the observed dipole being
purely due to our proper-motion and thus only consider harmonic modes with
$\ell \geq 2$ in the discussion below.
 
Harmonic techniques seem to be much better suited than angular-space methods
for extracting $\Lambda$CDM model parameters.
One reason is that for statistically isotropic skies
the harmonic coefficients are orthogonal in a statistical sense (i.e. uncorrelated) 
\begin{equation}
\langle a_{\ell m} a_{\ell' m'}^* \rangle = \delta_{\ell \ell'} \delta_{m m'} C_\ell.
\end{equation}
For Gaussian harmonic coefficients, all information is encoded in the angular power spectrum $C_\ell$.

In angular space, the two-point correlation  function of a statistically isotropic sky
\begin{equation}
C(\theta) \equiv \langle T(\unitvec{e}_1) T(\unitvec{e}_2) \rangle = \frac{1}{4 \pi} \sum_\ell (2 \ell + 1) C_\ell P_\ell (\cos\theta), 
\quad \unitvec{e}_1\cdot \unitvec{e}_2 = \cos \theta, 
\end{equation} 
does not have the property that $C(\theta)$ is independent from $C(\theta')$
for $\theta\neq\theta'$.  Thus it seems to be easier to draw inferences from
the angular band power spectrum than from the angular two-point correlation
function.  On the other hand, if a feature is attached to a certain region of
the sky, or otherwise violates statistical isotropy, it may be much harder to
spot it in the harmonic analysis than in angular-space.

The issue of cosmic variance is important for the analysis of the largest
cosmological scales.  For statistically isotropic and Gaussian skies, the
estimation of the angular power spectrum $C_\ell$ is limited by the fact that
we can only observe one particular realization of the Universe. For full sky
observations the estimator
\begin{equation}
\hat{C}_\ell = \frac{1}{2\ell +1} \sum_{m = - \ell}^{\ell} |a_{\ell m}|^2
\label{eq.eq:pseudo_Cl}
\end{equation}
is unbiased ($\langle \hat{C}_\ell \rangle = C_\ell$) and minimizes
variance,\footnote{Cosmic variance is defined for $\ell \geq 2$. For $\ell =0$
  it diverges. For $\ell =1$ it would be well defined, but cosmic variance
  does not apply if the CMB dipole is caused by the proper-motion of the Solar
  System.} 
\begin{equation}
  {\rm Var}[\hat{C}_\ell] = \frac{2}{2\ell +1} C_\ell^2.
  \label{eq:CV}
\end{equation}
The expression in Eq.~(\ref{eq:CV}) is the sample variance which, in
cosmology, is usually referred to as the cosmic variance: this is an
irreducible lower bound on the error in the measurements of the angular power
spectrum coming from the fact that we observe fluctuations in only one
universe.  The cosmic variance of the angular power spectrum also leads to a
nonzero cosmic variance of estimates of the two-point correlation
$C(\theta)$. The cosmic variance further increases when foreground dominated
regions are masked in the data analysis.  All results quoted below take these
aspects fully into account.

A summary of the most important findings in angular and harmonic space is
provided in Tab.~\ref{tab1}.

\subsection{Low variance and lack of correlation}

Historically, the first surprise, already within the COBE data, was the
smallness of the quadrupole moment.  When WMAP released its data \cite{WMAP1},
it confirmed $C_2$ to be low, however it was also shown that cosmic variance
allows for such a small value \cite{Efstathiou2003}.

\begin{table}
\hfill \begin{tabular}{|l|c|c|r|}
\hline
feature & p-value & data & reference \\
\hline
in angular space & & & \\
low variance ($N_{\rm side} = 16$) & $\leq 0.5\%$ &  Planck 15 & Tab.~12 \cite{Planck2015Isotropy} \\
2-pt correlation $\chi^2(\theta > 60^\circ)$ & $\leq 3.2\%$ & Planck 15 & Tab.~14 \cite{Planck2015Isotropy}  \\
2-pt correlation $S_{\rm 1/2}$ & $\leq 0.5 \%$ & Planck 15 & Tab.~13 \cite{Planck2015Isotropy}  \\
2-pt correlation $S_{\rm 1/2}$ & $\leq 0.3 \%$ & Planck 13 \& &  \\
                                                      &                          & WMAP 9yr & Tab.~2 \cite{CHSSctheta2015} \\
2-pt correlation $S_{\rm 1/2}$ (larger masks) & $\leq 0.1 \%$ & Planck13   & Tab.~2 \cite{CHSSctheta2015} \\
                                                                              & $\leq 0.1 \%$ & WMAP 9yr & \cite{CHSSctheta2015,Gruppuso2014}  \\
hemispherical variance asymmetry & $\leq 0.1 \%$ & Planck 15 & Tab.~20 \cite{Planck2015Isotropy} \\
cold spot & $\leq 1.0 \%$ & Planck 15 & Tab.~19 \cite{Planck2015Isotropy} \\
\hline
in harmonic space & & & \\
quadrupole-octopole alignment & $\leq 0.5 \%$ & Planck 13 & Tab.~7 \cite{CHSSalignments2015}\\
$\ell = 1,2,3$ alignment & $\leq 0.2 \%$ & Planck 13 & Tab.~7 \cite{CHSSalignments2015}\\
odd parity preference $\ell_{\rm max} = 28$ & $< 0.3\%$ & Planck 15 & Fig.~20 \cite{Planck2015Isotropy}\\
odd parity preference $\ell_{\rm max} < 50$ (LEE) & $ < 2\%$ & Planck 15 &  Text \cite{Planck2015Isotropy} \\
dipolar modulation for $\ell = 2$ -- $67$ & $\leq 1\%$ & Planck 15 & Text \cite{Planck2015Isotropy} \\
\hline 
\end{tabular}
\caption{P-values in per cent of various unexpected features. In this table we
  define the sense of p-values such that a small value means that it is
  unexpected. In some cases this is different from the sense used by the
  Planck collaboration in their analysis. The Planck analysis relies on just
  1000 Monte Carlo simulations of the instrument and pipeline and thus
  p-values below $0.2\%$ cannot be resolved. Other groups have used larger
  numbers of simulations, but those simulations do not include instrumental
  and algorithmic effects of the Planck analysis. LEE stands for look
  elsewhere effect. \label{tab1}}
\end{table}

\begin{figure}
\hfill \includegraphics[width=0.8\linewidth]{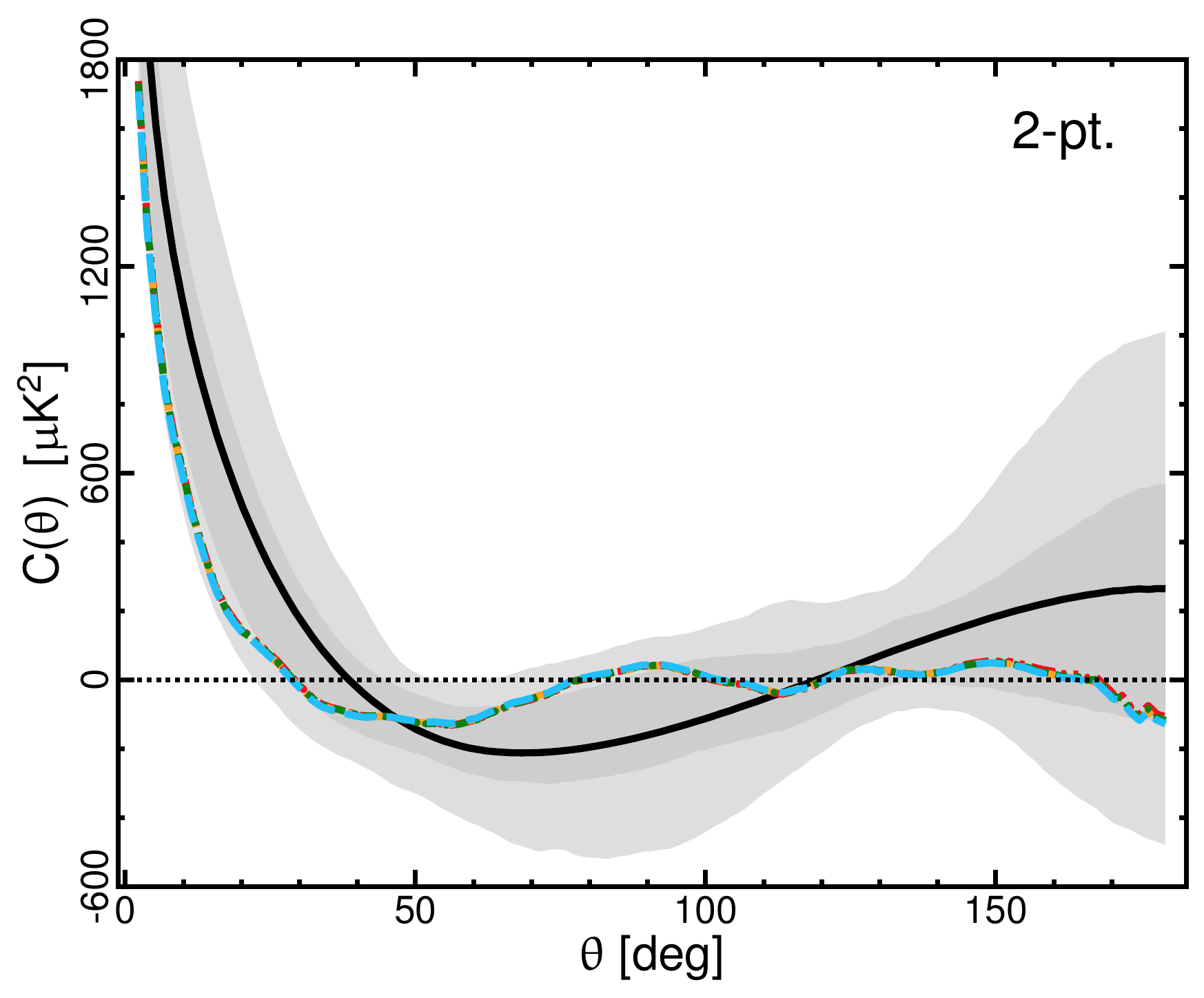}
\caption{Angular two-point correlation function as observed by Planck
  \cite{Planck2015Isotropy}. The full black line and the shaded regions are
  the expectation from 1000 SMICA simulations based on the $\Lambda$CDM model
  and the 68\% and 95\% confidence regions. The plot also shows four colored
  lines that fall on top of each other and represent the result of the Planck
  analysis of the Commander, SEVEM, NILC and SMICA maps at resolution $N_{\rm
    side} = 64$. While the measured two-point correlation is never outside the
  $95\%$ confidence region, the surprising feature is that we observe
  essentially no
  correlations at $70^\circ < \theta < 170^\circ$ and a significant lack of
  correlations at $\theta > 60^\circ$. \label{Ctheta}}
\end{figure}

Another rediscovery in the first release of WMAP \cite{WMAP1} was that the
angular two-point correlation function at angular scales $\gsim 60$ degrees is
unexpectedly close to zero, where a non-zero correlation signal was to be
expected.  This feature had already been observed by COBE
\cite{Hinshawetal1996}, but was forgotten by most of the community before its
rediscovery by WMAP.  The two-point correlation function as observed with
Planck \cite{Planck2015Isotropy} is shown in Fig.~\ref{Ctheta}.

The WMAP team suggested a very simple statistic \cite{WMAP1-cosmology} to characterize the vanishing correlation function --
\begin{equation}
S_{\mu} \equiv \int_{-1}^{\mu} {\rm d} (\cos \theta) [C(\theta)]^2,
\end{equation}
with $\mu \equiv \cos \theta = 1/2$. This measures the deviation from zero at
$\theta > 60^\circ$.  Detailed further investigations of the lack of angular
correlation have been presented in
\cite{CHSS2007,CHSS2009,WMAP7-anomalies,Gruppuso2014,CHSSctheta2015}.
Depending on the details of the analysis, p-values consistently below $0.5\%$
have been obtained, some even below $0.01 \%$.  An important question is the
size of the mask used in the analysis.  It has been shown in \cite{CHSS2009}
that most of the large-angle correlations in reconstructed sky maps are
between pairs of points at least one of which is in the part of the sky that
is most contaminated by the Galaxy.  This is in line with the findings of
\cite{Gruppuso2014}, where it was shown that more conservative masking makes
the lack of correlation even more significant.  This by itself already
signifies a violation of isotropy.

Undoubtedly, $S_{1/2}$ is an {\it ad hoc} and {\it a posteriori} statistic,
but it captures naturally the observed feature originally noted in COBE.
Several {\it a posteriori} ``improvements'' have been suggested
\cite{Hajian2007,Planck2015Isotropy}.  For example, in order to avoid the
argument that $\mu=1/2$ has been fixed after the fact one might let $\mu$
vary.  But now the look elsewhere effect must be taken into account.  The
Planck team implemented such an analysis which (in our convention) returns
global p-values of the order of $2\%$.  However, this global $S_\mu$ statistic
addresses a different question, namely how likely is it that there is a lack
of correlation for an arbitrary $\mu$. Thus we cannot argue that this
statistic is better than $S_{1/2}$, all we can say is that it is different.

Another critique was that the $S_{1/2}$ statistic does not account for
correlations among $C(\theta)$ at different $\theta$ \cite{Hajian2007}.  Such
a correlation is indeed expected in the $\Lambda$CDM model, but if we would
use that fact, we would be injecting a model assumption into the data
analysis.  Thus the recent Planck analysis \cite{Planck2015Isotropy} tests the
$\chi^2(\theta > 60^\circ)$ statistics, which compares the data to the
$\Lambda$CDM expectation, and a $\chi^2_0(\theta > 60^\circ)$ statistics,
which tests for the non-vanishing of $C(\theta)$ assuming the $\Lambda$CDM
covariance. The p-values for both tests are around $3$ and $2$ per cent. Let
us add that the recent Planck analysis is based on a resolution of $N_{\rm
  side} = 64$ and relies on a mask that includes $67\%$ of the sky for
cosmological analysis.  It was shown recently in \cite{Gruppuso2015} that
enlarging that mask gives rise to significantly smaller p-values and increased
evidence of a lack of correlation.

It has further been suggested \cite{Efstathiou:2009di} that 
the two-point correlation function calculated directly on a cut sky is a suboptimal
estimator  of the full-sky two-point correlation function, and that 
better estimators lead to less statistical significance for the observed
anomalies. This result has been extended \cite{Pontzen:2010yw} to all anisotropic
Gaussian theories with vanishing mean.  We however think that the issue of the
optimality of the full-sky estimator is irrelevant, since it is the {\it
cut-sky} two-point correlation function which is observed to be strikingly
anomalous, and which begs an explanation \cite{CHSSctheta2015}. 
% \dragan{[
% remove next sentence? not sure it's helpful]} That
% explanation may or may not involve an anomalous full-sky correlation function.
 
Another very simple statistic is to calculate variance, skewness and kurtosis
of the unmasked pixels.  In this test the Planck team found evidence for low
pixel variance for low-resolution maps ($N_{\rm side} = 16$), while skewness
and kurtosis behave as expected \cite{Planck2013Isotropy,Planck2015Isotropy}.
This feature seems to be consistent with a low quadrupole, a lack of power at
large angular scales (for $\ell < 30$, see Fig.~\ref{Planck2015_TT}) and the
discussed lack of angular correlation. The latter cannot be explained by a
lack of quadrupole power alone.  All modes below $\ell \leq 5$ contribute to
the observed lack of angular correlation \cite{CHSS2009} -- not by having low
amplitudes, but by combining to cancel one another and the contributions of
still higher $\ell$.  This is indicative of correlations among $C_\ell$ not
predicted by the $\Lambda$CDM model and a violation of both statistical
isotropy and scale invariance.

It has been noted \cite{CHSSctheta2015} that what is reported as the two-point
angular correlation function $C(\theta)$ is actually the dipole (and monopole)
subtracted two-point angular correlation function. A cosmological dipole of
the size expected in the best-fit $\Lambda$CDM model would completely dominate
$S_{1/2}$. In order not to raise $S_{1/2}$ significantly above its
dipole-subtracted value, one must have $C_1 \lesssim 200\mu K^2$, compared to
the value of approximately $3300\mu K^2$ that standard CMB codes return.  Of
course even the latter is orders of magnitude smaller than the dipole that we
measure, which we interpret to be entirely due to the observer's motion with
respect to the rest frame of the CMB (and hence excluded from $C(\theta)$).
There is disagreement over whether the intrinsic CMB dipole is physical or
observable.  It is certainly likely to be difficult to measure the dipole to
the necessary parts in $10^4$, to begin distinguishing the intrinsic dipole,
in any, from the $~3$mK Doppler dipole.

\subsection{Alignments of low multipole moments}

In the standard $\Lambda$CDM model the temperature (and other) anisotropies
have random phases.  In harmonic space this means that the orientations and
shapes of the multipole moments are uncorrelated.  This was first explored
in the first year WMAP data release using the angular momentum
dispersion~\cite{dOC-alignment} where it was discovered that the octopole
($\ell=3$) is somewhat planar (dominated by $m=\pm\ell$ for an appropriate 
choice of coordinate frame orientation)
with a p-value of about $5\%$. (A quadrupole
is always planar.)  More importantly, the normal to this plane (the axis
around which the angular momentum dispersion is maximized) was found to be
surprisingly well aligned with the normal to the quadrupole plane at a
p-value of about $1.5\%$.

\begin{figure}
  \hfill \includegraphics[width=0.8\linewidth]{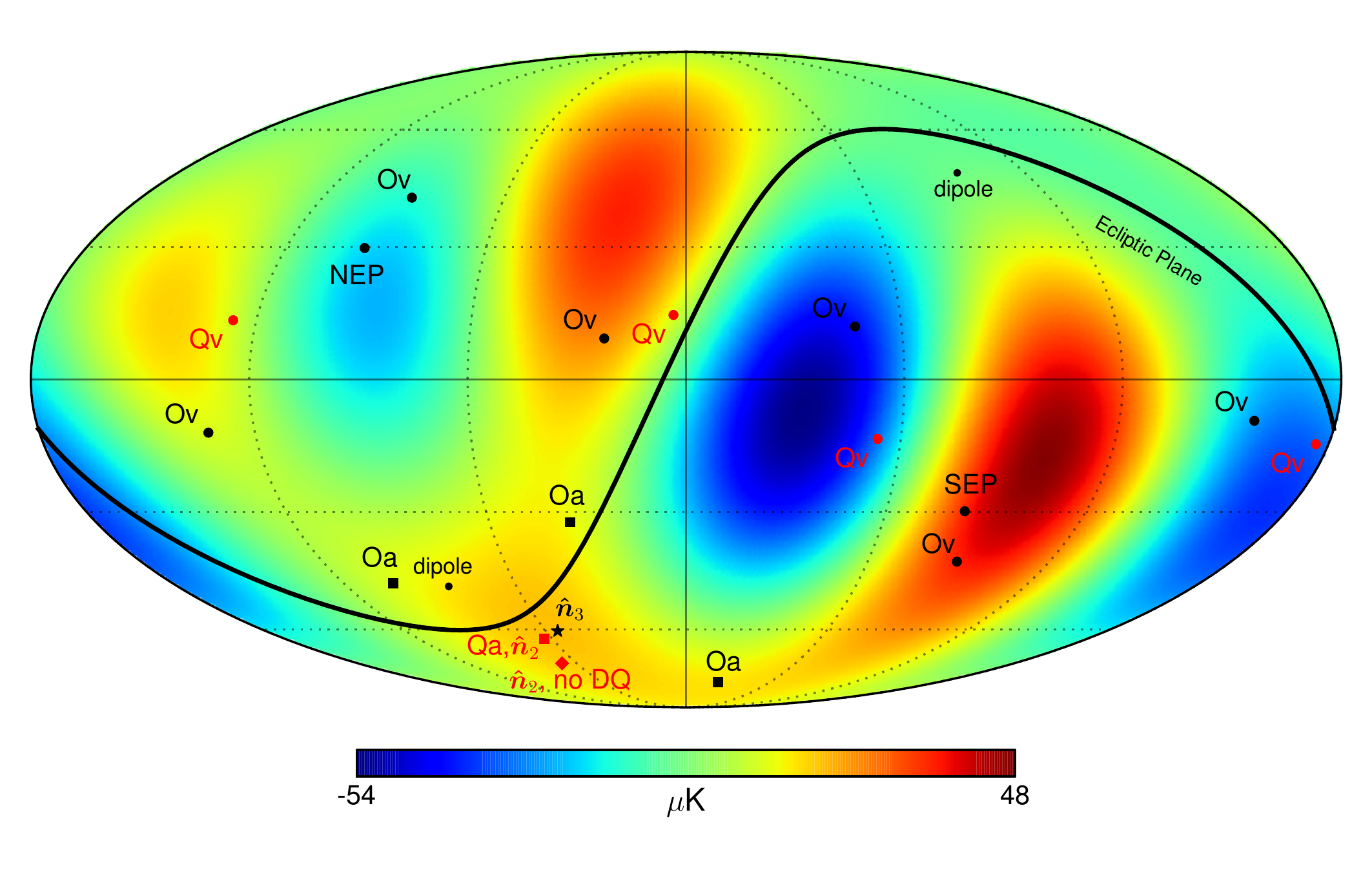}
  \caption{The combined quadrupole-octopole map from the Planck 2013
    release \cite{CHSSalignments2015}.  The multipole vectors (v) of the
    quadrupole (red) and for the octopole (black), as well as their
    corresponding area vectors (a) are shown. The effect of the correction
    for the kinetic quadrupole is shown as well, but just for the angular
    momentum vector $\hat{n}_2$, which moves towards the corresponding
    octopole angular momentum vector after correction for the understood
    kinematic effects.}
  \label{fig:MPV}
\end{figure} 

Subsequently these ideas have been studied in more detail using other measures
of planarity and alignment \cite{SSHC2004, Land-MPV, Land-axis,
  WMAP7-anomalies,CHSSalignments2015}.  A convenient tool for such a study are
the Maxwell multipole vectors~\cite{MPV}.  They provide an alternative to the
spherical harmonics as a means to represent angular momentum $\ell$ objects in
a manifestly symmetric, rotationally invariant manner.  Roughly speaking they
represent the multipole moments by products of unit vectors.  A dipole is a
vector.  A quadrupole can be constructed from the product of two dipoles (two
vectors), an octopole from three dipoles (three vectors), and so on for
arbitrary multipole moment $\ell$.  Of course two dipoles produce both
quadrupole and monopole moments so only particular combinations of the
products of dipoles will produce a pure quadrupole.  Mathematically they are
represented by the trace-free product of the two dipoles.  In the end this
means an angular moment $\ell$ object can be represented by $\ell$ unit
vectors and an overall amplitude (put together these contain the requisite
$2\ell+1$ degrees of freedom).

Given the multipole vectors, $\unitvec{v}^{(\ell;i)}$ for $i=1$ to $\ell$,
questions about alignments can now be addressed.  It has been found
convenient to directly study not the multipole vectors but instead their
oriented areas \cite{MPV}
\begin{equation}
  \vec w^{(\ell;i,j)} \equiv
  \unitvec{v}^{(\ell;i)}\times\unitvec{v}^{(\ell;j)},
\end{equation}
defined for each pair of multipole vectors at a fixed $\ell$.  Notice that
these are not unit vectors, their magnitudes are the area of the
parallelogram created by the two vectors.  These oriented-area vectors can
then be compared among the multipoles or to fixed directions.  The
multipole vectors for the quadrupole and octopole along with the oriented
area vectors, their maximal angular momentum dispersion directions $\unitvec{n}_\ell$,
and some special directions are shown in Fig.~\ref{fig:MPV}.

Numerous statistics can be defined to quantify alignment; here we only
consider one.  Since the multipole vectors really only define axes (both
$\pm\unitvec v^{(\ell;i)}$ are multipole vectors) we define the $S$
statistic as
\begin{equation}
  \label{eq:S-statistic}
  S \equiv \frac{1}{n} \sum_{j=1}^n | \vec w_j\cdot\unitvec e |.
\end{equation}
Here $\unitvec e$ represents a fixed direction on the sky and $\vec w_j$
represents one of the oriented-area vectors. The sum is over some set of
oriented-area vectors.  Although this can be used with any set of multipole
vectors and/or directions here we focus on two cases.  First, the
quadrupole-octopole alignment where we use $\vec w^{(2;1,2)}$ (the oriented
area vector for the quadrupole) as the fixed direction called $\unitvec e$
above and the $\vec w_j$ are the three oriented-area vectors for the octopole,
$\vec w^{(3;i,j)}$. Second, the joint alignment of the quadrupole and octopole
with special directions such as the normals to the Ecliptic and to the
Galactic planes or the direction of our motion with respect to the CMB (the
dipole direction).  In this latter case the $\vec w_j$ refer to the quadrupole
oriented-area vector and the three such vectors from the octopole (so that
$n=4$).

The most recent analysis of the latest WMAP and the Planck~2013 data
releases~\cite{CHSSalignments2015} finds the quadrupole and octopole
anomalously aligned with one another, with p-values ranging from about $0.2\%$ to $2\%$
depending on the exact map employed.  It is further found that the
quadrupole and octople are jointly perpendicular to the Ecliptic plane 
(i.e. their area vectors are nearly orthogonal to the normal to the Ecliptic) 
with a p-value of $2\%$ to $4\%$ and to the Galactic pole with a p-value of $0.8\%$
to $1.6\%$.  Even more strikingly they are aligned with the dipole
direction with a p-value of $0.09\%$ to $0.37\%$.

A number of issues must be considered when interpreting the p-values given above.  
Arguably, it is surprising not only that these alignments are observed at all
but that they have persisted in the data from the original
WMAP data release to the present.  To study the alignments (phase structure
of the temperature fluctuations) full-sky maps are required.  Thus the
results of \cite{CHSSalignments2015} are based on the cleaned maps produced by WMAP (the ILC maps) and
from \emph{different} cleaning methods employed by Planck (NILC, SEVEM, and SMICA).  
The exact phase structure of the maps are sensitive to
many effects including the details of the cleaning algorithms, systematics
effects in interpreting the data (such as the beam profile) which have been
improved through the years, and the different observation strategies
employed by WMAP and Planck.  Despite the many effects that could have masked the
alignments, they persist in the data and remain to be understood.

Our motion with respect to the rest frame of the CMB contributes not only to
the dipole, as mentioned above, but also to all other higher multipole
moments.  The effect of our motion in mixing two multipole moments $\ell$ and
$\ell' = \ell + \Delta \ell$ is suppressed by $\mathcal{O}(\beta^{|\Delta\ell|})$ with
$\beta\sim10^{-3}c$.  The monopole therefore contaminates mostly the dipole,
and has little effect on the power spectrum for $\ell>1$.  Other multipoles
mix only slightly, since they have comparable $C_\ell$ to begin with.
(Actually, the mixing effect is $\mathcal{O}((\beta\ell)^\ell)$, so there is
significant mixing at $\ell\simeq \beta^{-1}$, but we will not concern
ourselves here with such high $\ell$.)  However, the so called kinematic
quadrupole does affect the phase structure for $\ell=2$ \cite{SSHC2004}, as
seen in Fig.~\ref{fig:MPV}.  The direction of the quadrupole oriented-area
vector, labeled by Qa and $\unitvec n_2$ in the figure, shifts by about
$5^\circ$ from the ``no DQ'' value (diamond) to the corrected value (square)
when the kinematic quadrupole moment due to our motion is removed from the
full-sky temperature map prior to analysis.  The amplitude of the kinematic
quadrupole is frequency dependent and the cleaned, full-sky maps are
constructed from linear combinations of observations in many frequency bands
making the exact kinematic quadrupole calculation difficult to calculate.
This is exacerbated by calibration techniques which sometimes subtract some of
the frequency dependent kinematic quadrupole contribution.  The Planck~2013
data release provided estimates for the required correction
factor~\cite{Planck2013Isotropy} beyond the simple estimate used in the
results quoted above.  Interestingly when these corrections are applied the
alignment becomes even more anomalous.  For example, the p-value for the
alignment with the dipole direction drops to between $0.06\%$ and $0.23\%$
\cite{CHSSalignments2015}, or even less \cite{Notari:2015kla}.  The angular
momentum quadrupole-octopole alignment and the increase of the alignment due
to the kinetic effect has also been confirmed for the Planck~2015 data set
\cite{PhDFrejsel}.

In summary, the octopole is unexpectedly planar; the quadrupole and octopole
planes  are unexpected aligned with each other, and unexpectedly perpendicular to the
Ecliptic and aligned with the CMB dipole. These alignments have been robust in 
all full-sky data sets since WMAP's first release, and are found to be exacerbated
by proper removal of the kinematic quadrupole.  No systematics and no foregrounds
have been identified to explain these apparent violations  of statistic isotropy.

\begin{figure}
\hfill \includegraphics[width=0.48\linewidth]{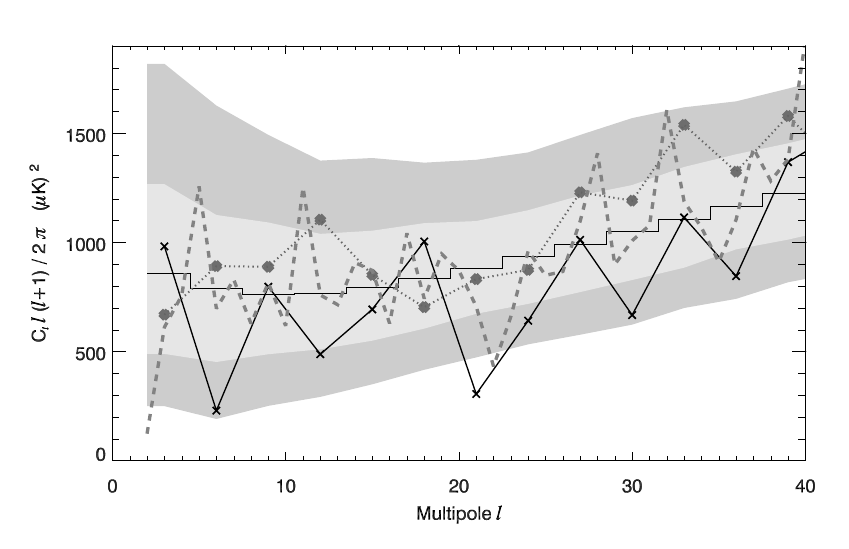}
\hfill \includegraphics[width=0.48\linewidth]{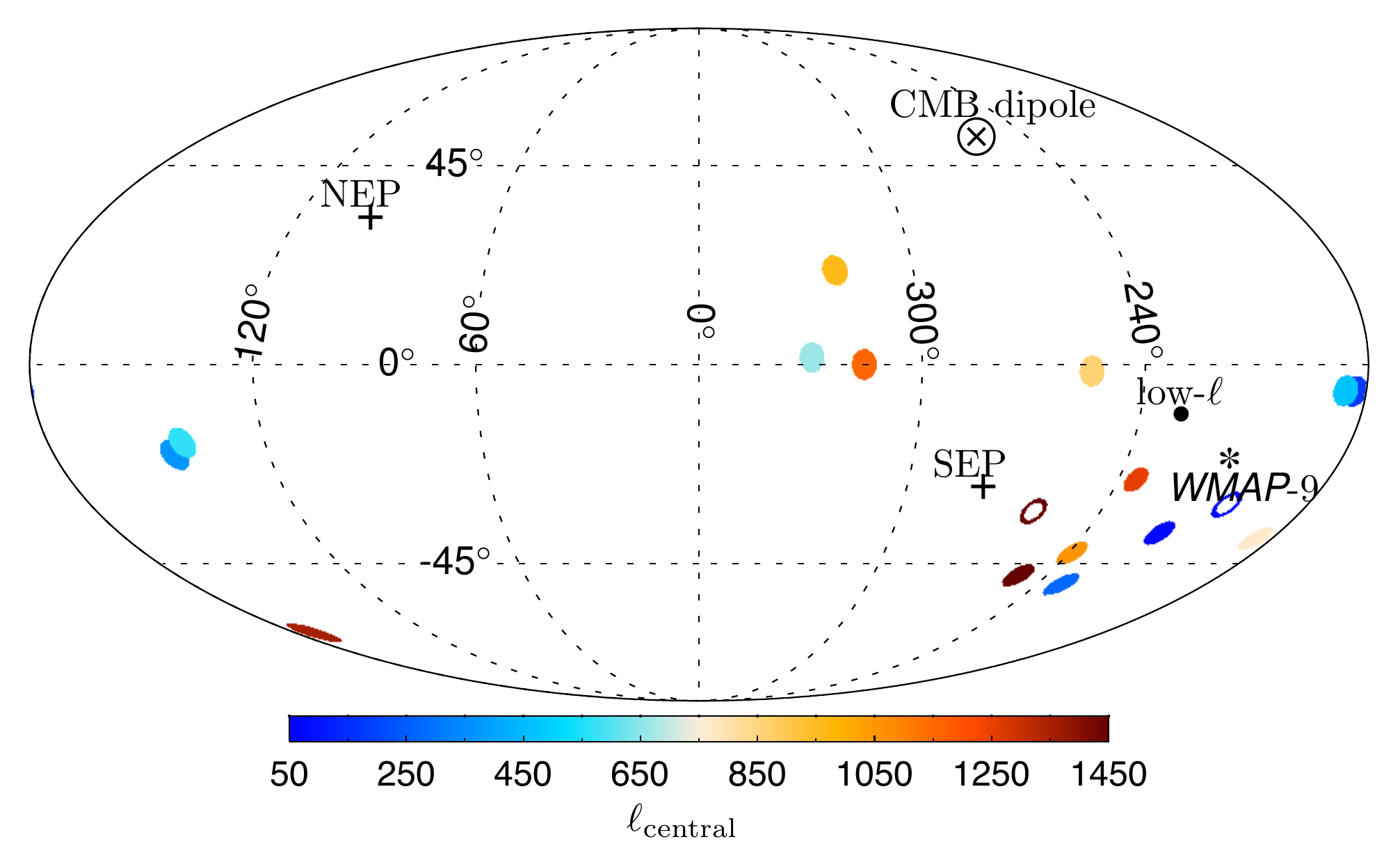}
\caption{Hemispherical power asymmetry. Left panel: original evidence, adopted
  from Ref.~\cite{Eriksen:2003db}. The three jagged lines show the binned
  angular power spectrum calculated over the whole unmasked sky (dashed),
  northern hemisphere (solid line, with crosses), and southern hemisphere
  (dotted line, with circles). North and south were defined with respect to
  the best-fit axis for WMAP1 data, and were close (but not identical) to the
  north and south ecliptic.  The histogram and the two grey areas around it
  denote the mean and the 68\% and 95\% confidence regions from Gaussian
  random simulations. Right panel: best-fit directions from the dipolar
  modulation model, applied to Planck 2015 SMICA map, evaluated in multipole
  bins centered at 50 to 1450 \cite{Planck2015Isotropy}. Directions
  corresponding to the North Ecliptic Pole (NEP) and South Ecliptic Pole
  (SEP), the CMB dipole, and the best-fit WMAP9 modulation direction are also
  shown. The ``low-l'' direction refers to constraining $\ellmax=600$, while
  the blue and brown rings show analysis in the two multipole ranges $\ell\in
  [2, 300]$ and $\ell\in [750, 1500]$, respectively.}
\label{fig:hemi_asym}
\end{figure}

\subsection{Hemispherical Asymmetry}

Evidence for hemispherical power asymmetry first emerged in the analysis of
WMAP first-year data \cite{Eriksen:2003db,Hansen:2004mj}. It was found that
the power in discs on the sky of radius $\sim 10^\circ-20^\circ$, evaluated
in several multipole bins, is larger in one hemisphere on the sky than the
other; see the left panel of Fig.~\ref{fig:hemi_asym}. The plane that
maximizes the asymmetry is approximately the Ecliptic, though it depends
somewhat on the multipole range; the variation of the normal to this plane
with multipole range is shown in the right panel of Fig.~\ref{fig:hemi_asym}.
Fig.~\ref{fig:MPV} shows that the combined 
quadrupole and octopole moment already contribute to such a power asymmetry.

\begin{figure}
\hfill \includegraphics[width=0.45\linewidth]{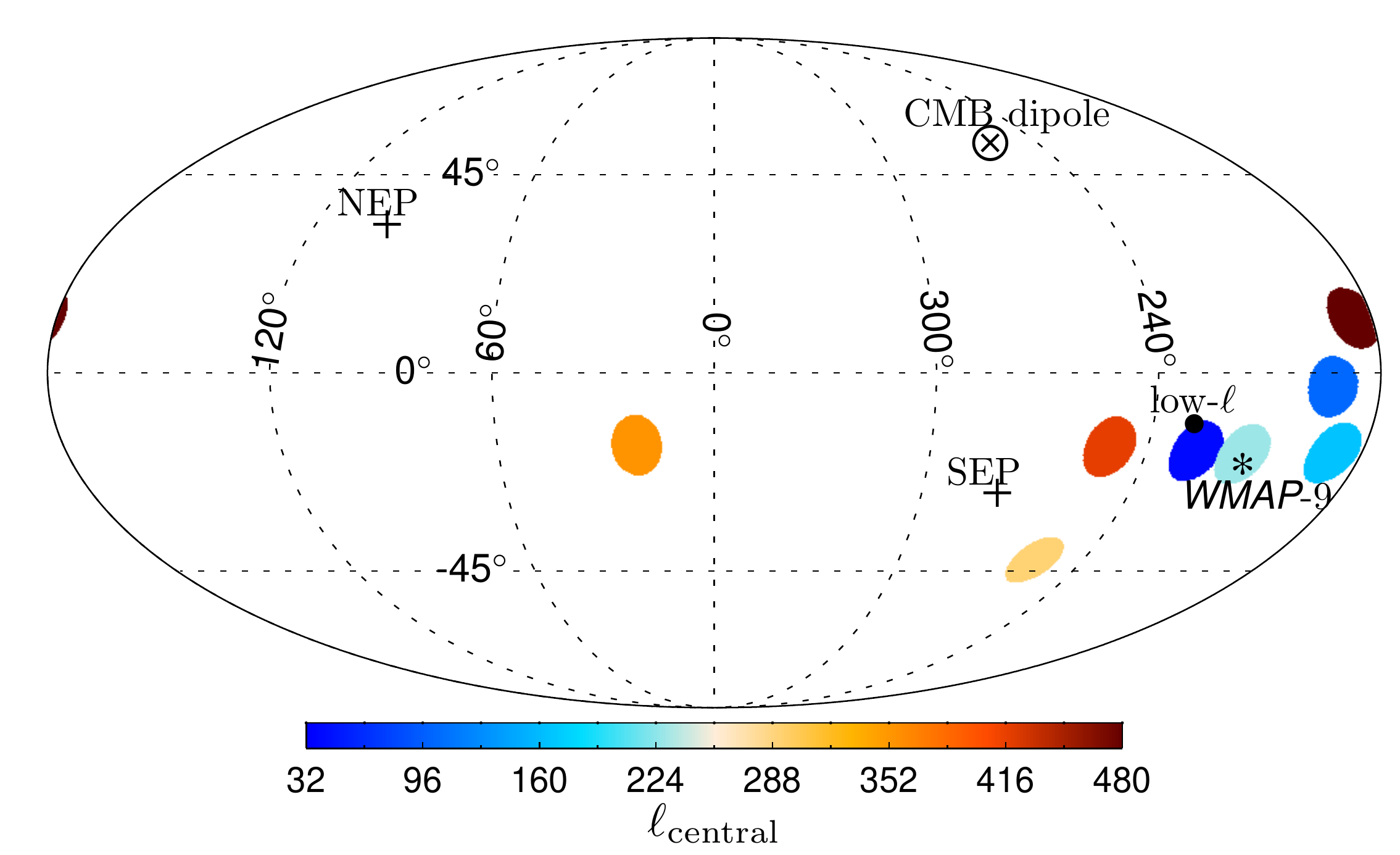}
\hfill \includegraphics[width=0.45\linewidth]{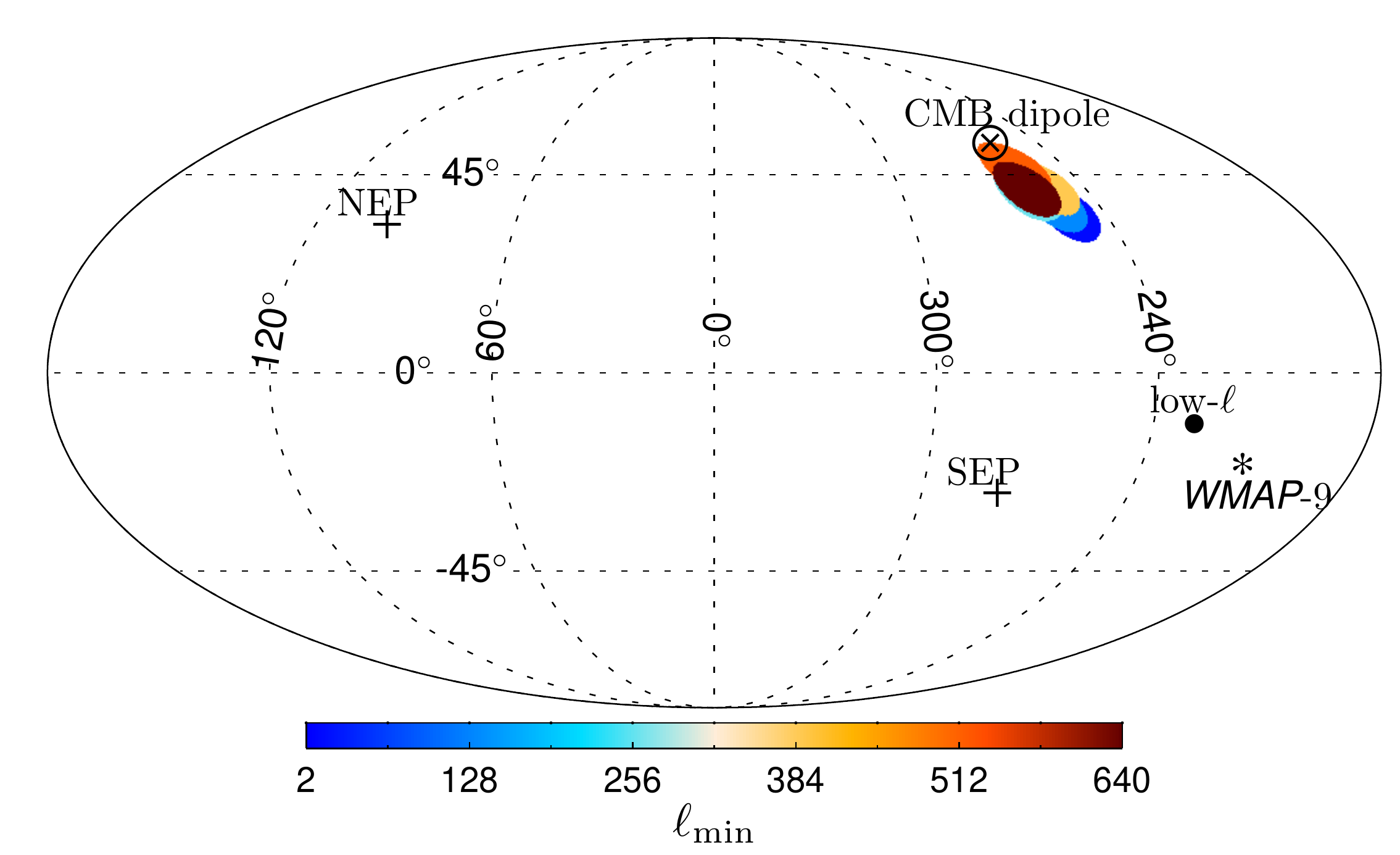}
\caption{Results of bipolar spherical harmonics analysis
  \cite{Planck2015Isotropy}. Left: Dipolar modulation, only darkest blue spot
  is statistically significant ($\ell = 2$ to $64$). Right: Doppler modulation
  at high $\ell$. \label{DM}}
\end{figure}

The study of hemispherical asymmetry was extended  to later years of WMAP
\cite{Eriksen:2007pc,Hoftuft:2009rq,Akrami:2014eta} as well as Planck
\cite{Akrami:2014eta,Planck2013Isotropy,Planck2015Isotropy} by analyses that
modeled the asymmetry as a dipolar modulation \cite{Gordon2005,Prunet:2004zy}
\begin{equation}
  T(\unitvec{e}) = T_0(\unitvec{e})[1+ A\,\unitvec{e}\cdot\unitvec{d}]
  \label{eq:dipole_mod}
\end{equation}
where $T(\unitvec{e})$ and $T_0(\unitvec{e})$ are the modulated and
unmodulated temperature fields, respectively, $\unitvec{e}$ is an arbitrary
direction on the sky, and $A$ and $\unitvec{d}$ are the dipolar modulation
amplitude and direction.  This parameterization enables a straightforward
Bayesian statistical analysis. The earlier analyses have found statistically
significant evidence for $A\sim 0.1$, and direction $\unitvec{d}$ roughly in
the ecliptic pole direction. The result from the Planck 2015 release, using
the Commander map, is $A=(0.066\pm 0.021)$ with $\unitvec{ d}$ pointing in the
direction $(l, b) = (230^\circ, -16^\circ)\pm 24^\circ$
\cite{Planck2015Isotropy}. The modulation's direction is remarkably consistent
as a function of the multipole range used, and between WMAP and Planck, as the
right panel of Fig.~\ref{fig:hemi_asym} shows. Planck also finds that the
modulation, as measured by the coupling of adjacent multipoles, has most
signal at relatively low multipoles, $\ell\in [2, 67]$ where it has a p-value
of $1\%$. In Fig.~\ref{DM} the dipolar directions found in Planck 2015 data
\cite{Planck2015Isotropy} by means of a bipolar spherical harmonics analysis
\cite{BipolSH1,BipolSH2} are shown. At low multipoles the same direction in
the Southern ecliptic hemisphere is identified (left panel), while at much
higher multipoles the Doppler boost and aberration dipolar modulation (due to
the proper motion of the Solar System) is picked up (right panel).

A different way to measure the hemispherical asymmetry is to consider variance
calculated on hemispheres. Refs.~\cite{Akrami:2014eta,Adhikari:2014mua} found that the
northern hemisphere in WMAP 9-year and Planck 2013 maps has an extremely low
(significant at $3$--$4\sigma$) variance evaluated on scales
$4^\circ$--$14^\circ$ relative to what is expected in the $\Lambda$CDM model. Planck
\cite{Planck2015Isotropy} found that the result holds at an even wider range
of scales once the lowest harmonics ($\ell\lesssim 5$) are filtered out
from the map.  Moreover, configurations of the three and four-point
correlation function, evaluated at a resolution of $\Nside=64$ (that is, down
to $\sim 0.5^\circ$ on the sky) also exhibit the hemispherical
asymmetry. Finally, evidence for hemispherical asymmetry in WMAP data was also
found by measuring power in disks of fixed size on the sky \cite{Hansen:2008ym}.

The fact that the axis that maximizes the asymmetry is close to the ecliptic
pole motivates both systematic and cosmological proposals for the
hemispherical asymmetry. Nevertheless, there have been no convincing proposals
to date about why one ecliptic hemisphere should have less power than the
other. 

\subsection{Parity asymmetry}

It is interesting to ask whether the CMB sky is, on average, symmetric with
respect to reflections around the origin, $\unitvec{e}  \rightarrow -\unitvec{e}$. 
The standard theory does not predict any particular behavior with
respect to this point-parity symmetry. Because $\Ylm (-\unitvec{e})
= (-1)^\ell \Ylm (\unitvec{e})$, even (odd) multipoles $\ell$ have an even
(odd) symmetry.

\begin{SCfigure}[][!t] % caption next to figure
%\begin{figure}
\hfill \includegraphics[width=0.55\linewidth]{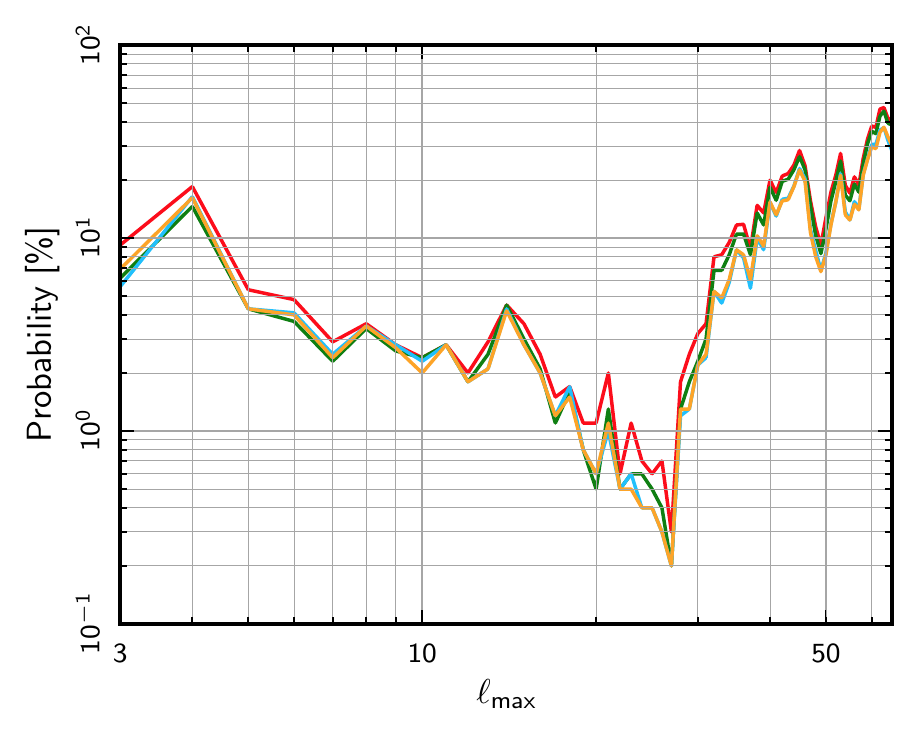}
\caption{Parity asymmetry in the Planck 2015 data \cite{Planck2015Isotropy}.  
  Shown is the p-value significance, based on a power-spectrum statistic sensitive to 
  parity for the four foreground cleaned Planck maps (Commander, NILC, SEVEM and SMICA) 
  as a function of the maximum multipole used in the analysis.}
\label{fig:parity}
\end{SCfigure}

Tests of parity of the CMB have first been discussed in
Ref.~\cite{Land:2005jq}, who studied both the aforementioned point-parity
symmetry, and the mirror parity ($\unitvec{e} \rightarrow \unitvec{e} -
(\unitvec{e} \cdot \unitvec{m})\unitvec{m}$, with $\unitvec{m}$ being the axis
normal to the mirror plane). The point-parity symmetry analysis of WMAP
maps was extended by \cite{Kim:2010gf,Kim:2010gd,Kim2011} who studied the even and odd
parity maps,
\begin{equation}
  T^+(\unitvec{e}) = \frac{T(\unitvec{e}) + T(-\unitvec{e})}{2}, \qquad
  T^-(\unitvec{e}) = \frac{T(\unitvec{e}) - T(-\unitvec{e})}{2}.
\end{equation}
Using a suitably defined power spectrum statistic -- the ratio of the sum over
multipoles of $D_\ell$ for the even map to that for the odd map -- they found
a $99.7\%$ evidence for the violation of parity in WMAP7 data in the multipole
range $2\leq\ell\leq 22$. The analysis was finally extended to Planck by
\cite{Planck2013Isotropy,Planck2015Isotropy}, who confirmed the results from
\cite{Kim:2010gf} based on WMAP, but also found that the significance depends
on the maximum multipole chosen, and peaks for $\ellmax\simeq 20-30$, but is
lower for other values of the maximum multipole used in the analysis; see
Fig.~\ref{fig:parity}. The corresponding p-values of the Planck 2015 analysis,
also including the `look elsewhere' effect with respect to the choice of
$\ell_{\rm max}$, are reported in Tab.~\ref{tab1}. Planck also studied the
mirror symmetry, finding less anomalous results than those for the
point-parity symmetry \cite{Planck2015Isotropy}.

%\end{figure}

In \cite{Naselsky:2011jp} it was shown for WMAP 7-years data 
that the directions of maximal (minimal) parity asymmetry for multipole 
moments up to $\ell \sim 20$, and excluding the $m=0$ modes from the analysis, 
seem to be normal (parallel) to the direction singled out by the CMB dipole. 
The direction that maximizes this parity asymmetry is also close to the direction of 
hemispherical asymmetry when including the lowest multipole moments. Thus 
parity asymmetry and hemispherical asymmetry might be linked to each other.

Whether the observed parity asymmetry is a fluke, an independent anomaly, or a
byproduct of another anomaly, is not clear at this time. The parity asymmetry
appears to be correlated with the missing power at large angular scales, as
the wiggles in the lowest multipoles, seen clearly in the top panel of
Fig.~\ref{fig:parity}, combine to nearly perfectly cancel the angular
two-point correlation function above 60 degrees \cite{CHSS2009}.

\subsection{Special regions: the cold spot} 

\begin{figure}
\includegraphics[width=0.45\linewidth]{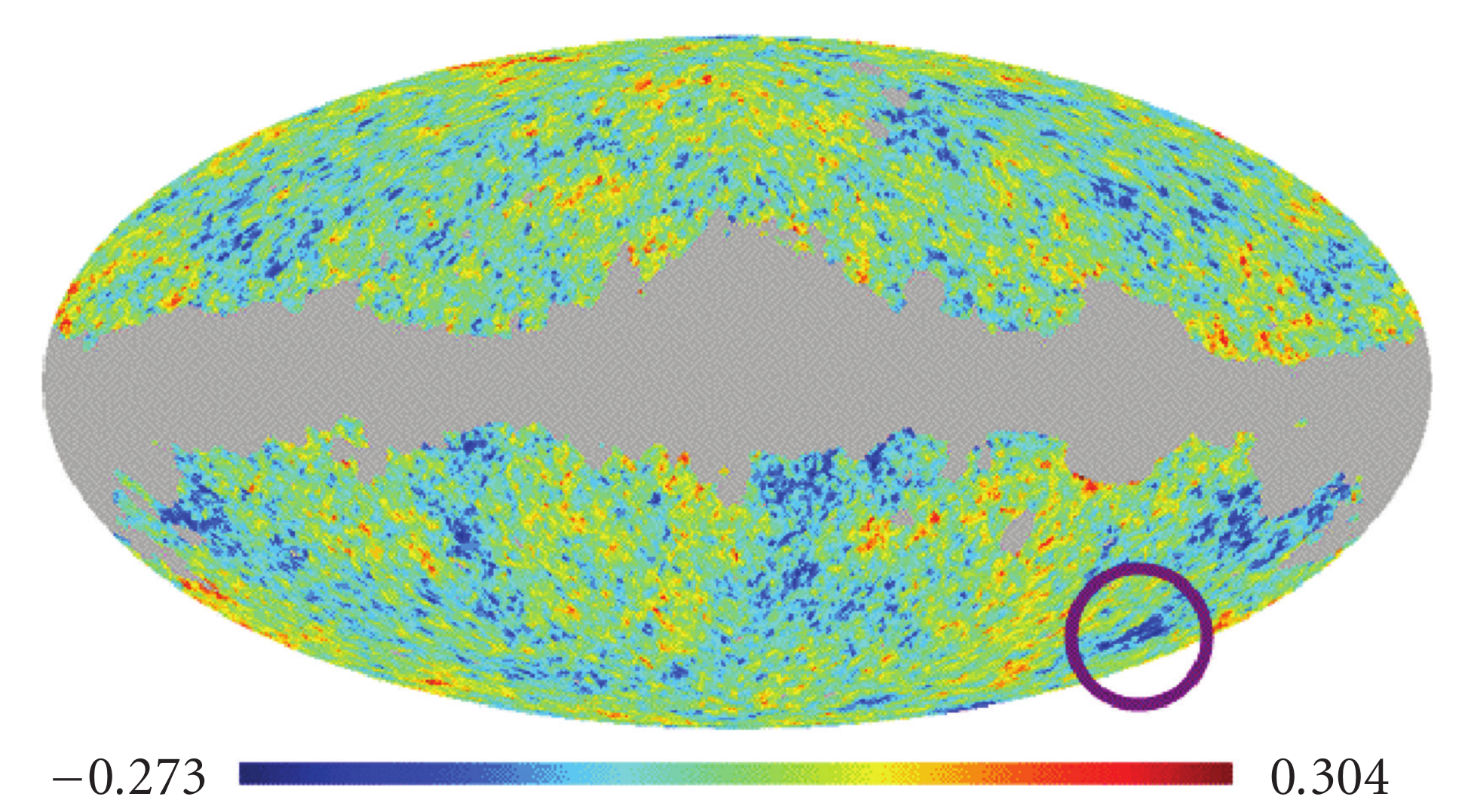}
\hspace{0.2cm} \includegraphics[width=0.45\linewidth]{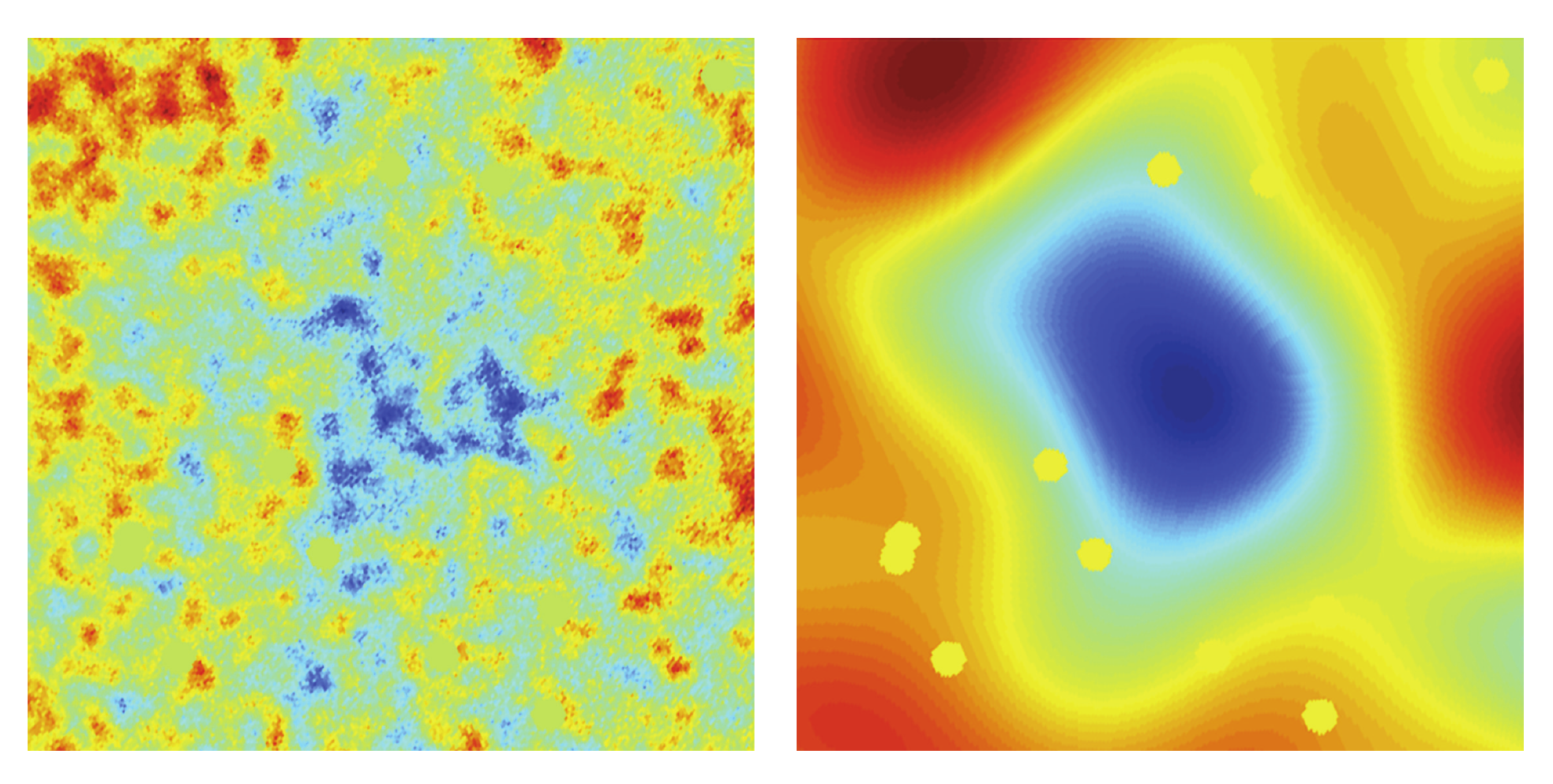}
\caption{Cold spot in WMAP 7th year temperature maps. Left panel shows the map
  with the circle. Middle panel is the more detailed picture of the spot,
  while the right panel is the wavelet-filtered version of the middle panel
  (wavelet size $R=250'$). The small spots in the right panel are regions of
  known point sources that have been masked). All figures are adopted from
  the review in Ref.~\cite{Vielva:2010ng}. \label{CS}}
\end{figure}

Evidence for an unusually cold spot in WMAP 1st year data was first presented
in \cite{Vielva:2003et}. The spot, shown in Fig.~\ref{CS}, is centered
on angular coordinates $(l, b) = (207^\circ, -57^\circ)$, has a radius of
approximately five degrees, is roughly circular \cite{Cruz:2006sv}, and the
evidence for its existence is frequency independent \cite{Vielva:2003et}. The
cold spot was originally detected using spherical Mexican hat wavelets, which are
well suited for searching for compact features on the sky; tests in
\cite{Vielva:2003et} detected a deviation from the Gaussian expectation in the
kurtosis of the wavelet coefficients at the wavelet scale of
$R=300'$.  Taking into account the `look elsewhere' effect, that is the fact
that not all statistics attempted with the wavelets returned an anomalous
result, \cite{Cruz:2006fy} estimated the statistical level of anomaly of the
cold spot to be $1.85\%$.

These original detections were followed up, confirmed, and further
investigated using not only Mexican hat wavelets
\cite{Cruz:2004ce,Chiang2007}, but also steerable \cite{Vielva:2007kt}
and directional \cite{McEwen:2004sv} wavelets, needlets
\cite{Rath:2007ti,Pietrobon:2008rf}, scaling indices \cite{Rossmanith:2009cy},
and other estimators \cite{Cayon:2005er}.  The detection of the cold spot has
also been challenged by \cite{Zhang:2009qg} on grounds that alternative
statistics -- say, over/under density at disks of varying radius -- does not
lead to a statistically significant detection once look-elsewhere effects are
taken into account. While the lingering worries about {\it a posteriori} nature of
this particular anomaly make its significance difficult to quantify, the
basic existence of the cold spot seems to be confirmed by most analyses. For
comprehensive overviews of the cold spot, see
\cite{Cruz:2009nd,Vielva:2010ng}.

The intermediate size of the cold spot, as well as its frequency independence,
argue against simplest systematic and foreground explanations. The size of the
cold spot ($\sim 10^\circ$) makes it too large to be a point source, yet
typically too small to be a diffuse foreground, especially since it is found
in a relatively foreground-clean part of the sky. And while the
Sunyaev-Zeldovich effect -- inverse Compton scattering of the CMB photons off
hot electrons in galaxy clusters -- could in principle lead to the desired
amplitude and spatial extent of the signal, the SZ effect has a very
pronounced frequency dependence that is completely incompatible with the
observed frequency independence of the cold spot signal \cite{Cruz:2006sv}.

Recent developments in the search for links between CMB cold spots and
underdensities in the galaxy distribution, discussed further in
Sec.~\ref{sec:LSS}, are of particular interest. While it is in principle
possible that an underdensity in the galaxy and dark matter distribution be
responsible for the CMB cold spot \cite{Inoue2006}, such a void would have to
be huge, and therefore fantastically unlikely in the standard $\Lambda$CDM
cosmology, making it much less probable than the CMB cold spot itself. One
could nevertheless search for the link between the CMB cold/hot spots and
galaxy under/overdensities. The most general way to search for such a link is
to cross-correlate the CMB temperature with the galaxy overdensity over the
whole observed sky (for each), but such tests have not shown evidence for
departures from the $\Lambda$CDM prediction
(e.g.\ \cite{Giannantonio:2012aa}). However, it is possible that the
cross-correlation performed more selectively -- e.g. looking for CMB
overdensity behind clusters of galaxies or voids
\cite{Granett:2008ju,Kovacs:2015bda} or, taken to extreme, behind the cold
spot alone \cite{Granett:2009aw,Bremer:2010jn} -- would indeed show departures
from $\Lambda$CDM predictions. Such tests performed to date have shown
tantalizing, though as yet not definitive, evidence for a large underdensity
in the distribution of galaxies in the same direction as the cold spot; this
is further discussed in Sec.~\ref{sec:local_LSS}. 

If the cold spot is indeed taken as a sign of departure from the $\Lambda$CDM
model's predictions, it may be possible to explain it using novel theory.  A
theoretical explanation has a challenge of generating a localized feature of a
rather small size ($\sim 10^\circ$) in a non-special direction on the sky. In
this regard, Bianchi cosmological models that are homogeneous but not
isotropic   are well suited and have been proposed as the explanation of the
cold spot \cite{Bridges:2007ne}. Another possibility that has been discussed
is the presence of cosmic textures \cite{Cruz:2007pe,Cruz:2008sb}, defects
whose profile parameters can be chosen to explain the cold spot. While the
texture explanation is favored by the Bayesian analysis \cite{Cruz:2007pe} and
appears viable in principle, it seems difficult to make further progress
without independent predictions made by the texture model and their
confirmation with future data.

\subsection{Special regions: loop A}

In the context of the study of a possible foreground from radio loops
(see Sec.~\ref{foregrounds}), a huge loop, named loop A, has been identified
in the vicinity of the cold spot \cite{PhDFrejsel}.  Masking this particular
region of the sky reduces the significance of the parity asymmetry, and the
significance of dipolar modulation, and this region might largely be
responsible for the observed quadrupole-octopole pattern.  However, a
corresponding foreground has not yet been identified.

\subsection{Statistical independence of CMB anomalies}

Although most of the described features or anomalies show p-values in the per
cent or per mille level, none of them individually reaches the $5\sigma$
detection level that is adopted in particle physics. Whether such a strong
criterion is actually necessary or not might be debated. However, it is
extremely hard to believe that our realization of $\Lambda$CDM just happens to
have all of these features by chance, unless they have a common origin.  It
might be that some of the features result in other features, e.g. a low
quadrupole clearly contributes to the low variance and {\it vice versa}.  Thus
in order to better characterize the CMB anomalies it would be useful to reduce
them to a few ``atoms'', i.e. a set of mutually independent features when
analyzed in the context to the inflationary $\Lambda$CDM model.  This study is
an ongoing program, and is computationally expensive, as it requires large
sets of Monte Carlo studies (many more than produced for the Planck Full Focal
Plane Simulations \cite{Planck2015Sim}).

Here we propose such a set of ``atoms'' that are independent of each other in the context
of the $\Lambda$CDM model.  At the present time we think that there are at least three such
``atoms'': lack of angular correlation at large angles, alignments of the
lowest multipole moments, and hemispherical power asymmetry.

The first ``atom'', the lack of correlation, can also cause a low quadrupole and 
low variance, while for example a low quadrupole alone, cannot cause a lack of 
correlation. Detailed studies of constrained simulations have shown that a lack of
correlation does not increase the chances to find alignments, and aligned
multipoles do not increase the probability to find a lack of correlation
\cite{Rakic2007,SHCSS2011}. It has also been investigated if an intrinsic
alignment of quadrupole and octopole correlates with the extrinsic alignment
with some other directions, such as the dipole, ecliptic or galactic planes.
These tests have been inconclusive \cite{CHSSalignments2015}. We thus 
conclude that the mutual alignment of the lowest multipole moments is the 
second anomaly ``atom''.

The observed dipolar modulation seems to be also independent of the alignments
between multipole moments.  We propose that to be the third ``atom''.
In \cite{Polastrietal2015} the p-values for
various alignment measures were compared to the predictions of $\Lambda$CDM and
a dipolar modulated model \cite{Gordon2005}. They showed that in both cases
the alignment p-values are of the order of $0.1\%$.  To our knowledge an
explicit test in which the lack of angular correlation is correlated with
dipolar modulation has never been done. The Planck team has also shown that a
high-pass filter, which suppresses the multipole moments at $\ell \leq 5$
actually increases the significance of the hemispherical variance asymmetry
\cite{Planck2015Isotropy}.  This shows that the
quadrupole-octopole pattern alone is not responsible for most of the
  hemispherical asymmetry signal; moreover, the maximal asymmetric directions for 
$\ell < 5$ and $\ell > 5$ do not agree, 
which is yet another indication that we face at
least three independent ``anomaly atoms''.

One of those ``atoms'' could certainly be an unlikely statistical fluke, however, it is
quite unlikely that two of them, or even all three of them are statistical
flukes (the corresponding p-values would be at most $\sim 10^{-5}$).  This
means that these anomalies signal a significant discovery of some new CMB
features.  The open question is, which of those features hold the key to
decipher the underlying physical mechanism(s).

\section{Foregrounds \label{foregrounds}} 

If these anomalous CMB features are related to local physics, it might not be
surprising that they appear to be a rare fluke.  The reason is simple -- our
environment is one particular example of an environment for a CMB mission and
every particular realization is somewhat special.  In this section we review
some of the local physical effects that have been suggested as explanations
for CMB anomalies.  We ignore speculations on instrumental effects, as it
seems to us that the consistency of WMAP and Planck 2015 results
\cite{Planck2015Overview} makes such an explanation quite unlikely.

\begin{figure}
\hfill \includegraphics[width=0.9\linewidth]{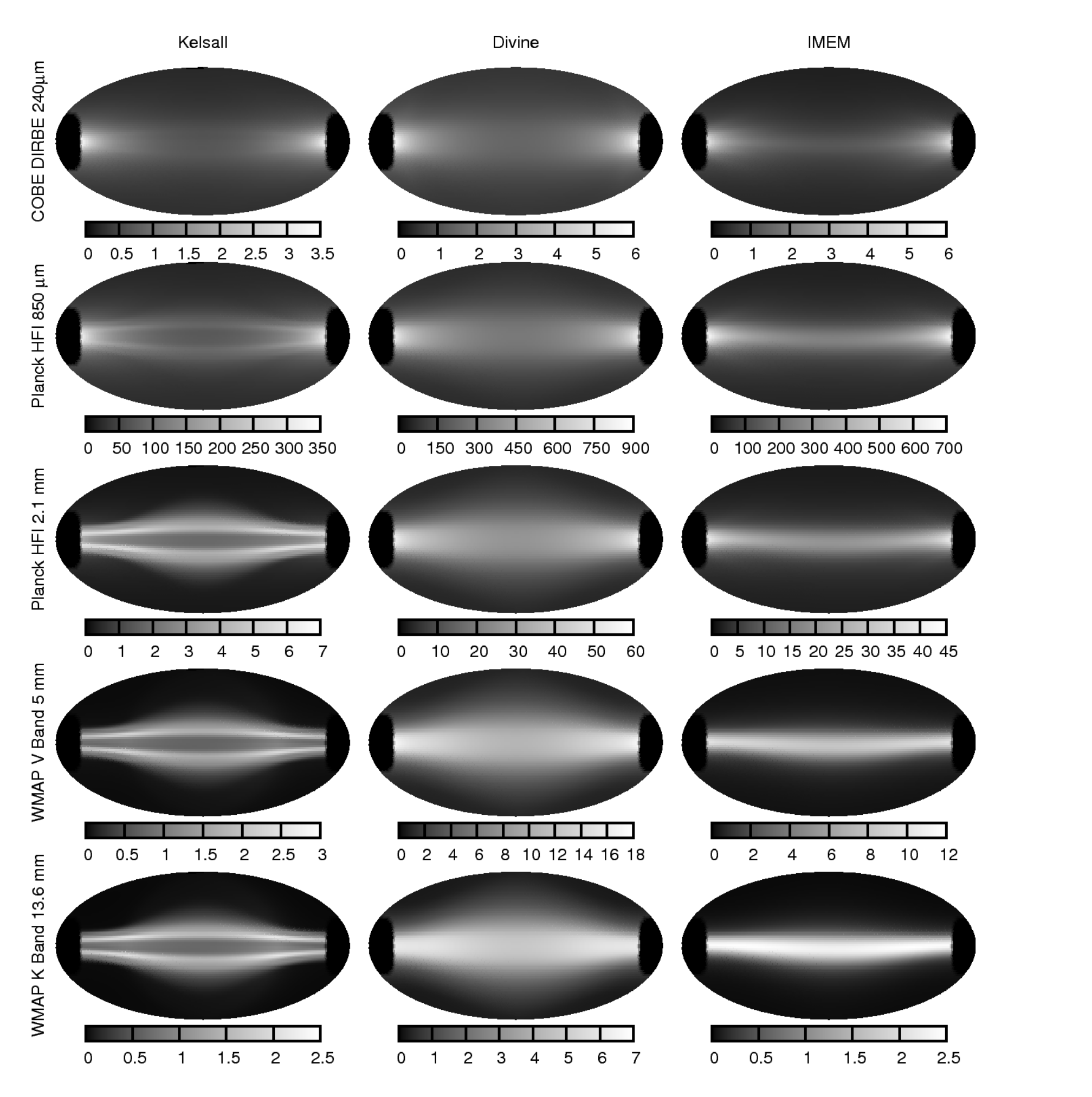}
\caption{Comparison of three contemporary models of Solar System dust
  \cite{DikarevSchwarz2015}.  The plots show all-sky maps of the thermal
  emission from the Zodiacal cloud as seen from Earth at the fall equinox
  time. The maps are in ecliptic coordinates and are centered on the vernal
  equinox. A disk of radius $30^\circ$ around the Sun is masked. The Planck
  analysis is based on the Kelsall model (left column). Note the much
  higher expected fluxes in the IMEM and Divine model (middle and
    right column), which assume here that the dust grains are carboneous.
  The grey scale of the upper row of maps is in MJy sterad$^{-1}$, the other
  rows are in $\mu$K of a temperature in excess of the CMB. Each map has its
  own brightness scale. \label{DustModels}}
\end{figure} 

\subsection{Solar System}

The closest foreground to a CMB space mission is the Solar System itself. An
obvious (subdominant) source of microwave radiation is the dust grains, and their
emission might contribute to or modify the observed CMB anomalies
\cite{Dikarev2008,Dikarev2009,Hansen2012}. The zodiacal cloud has been studied
in detail for the Planck 2013 release \cite{Planck2013ZD} and the Planck team
in its 2015 analysis subtracted a fit to the Kelsall model for the zodiacal
cloud before map making \cite{Planck2015Overview}. The Kelsall model \cite{Kelsall} 
attempts to capture the Solar System dust emission in the infrared and microwaves and 
is based on the analysis of COBE DIRBE observations.

When comparing the Kelsall model with two meteoroid engineering models (used
by space agencies to reduce the hazard to launch a spacecraft into a shower of
meteoroids), it has been found that those engineering models
\cite{Divine,Dikarev}, depending on the chemical composition of the dust
grains, predict a much brighter zodiacal cloud at microwave frequencies
\cite{DikarevSchwarz2015}. The Divine model of the interplanetary meteoroid
environment \cite{Divine} predicts meteoroid fluxes on spacecraft anywhere in
the Solar system from $0.05$ to $40$~AU from the Sun. This model uses data
from micro-crater counts in lunar rocks from Apollo, meteor radar, and in situ
measurements from Helios, Pioneer 10 \& 11, Galileo and Ulysses. However, it
does not make use of the infrared observations of COBE DIRBE.  The
Interplanetary Meteoroid Engineering Model (IMEM) \cite{Dikarev} and the
Divine model use the distributions in orbital elements and mass rather than
the spatial density functions of the Kelsall model, ensuring that the dust
densities and fluxes are predicted in accord with Keplerian dynamics of the
constituent particles in heliocentric orbits. IMEM is constrained by the
micro-crater size statistics collected from the lunar rocks, COBE DIRBE
observations of the infrared emission from the interplanetary dust at 4.9, 12,
25, 60, and 100~$\mu$m wavelengths, and Galileo and Ulysses in-situ flux
measurements.

The different predictions for microwave emission from the Solar System of the three models 
is illustrated in Fig.~\ref{DustModels}. 
%\dragan{[is next sentence important
%    to keep?]} The Planck fit to the Kelsall model returns negative
%absorption cross sections for several components (most importantly the main
%cloud at $100$~GHz), which are unphysical and demonstrate that this model still
%needs to be improved. 

To a first approximation, the zodiacal dust foreground produces a smooth band
along the ecliptic, see Fig.~\ref{DustModels}. This does not give rise to a
hemispherical asymmetry, but it could cause alignments of low CMB multipoles
with the ecliptic plane.  Additionally, it could contribute to a positive
correlation at very large angular separations, as the antipode of a point
close to the ecliptic is also close to the ecliptic.  However, the shape of
the emission from the zodiacal cloud cannot give rise to the type of alignment
observed in the low $\ell$ multipoles of the CMB because it looks
looks like a $Y_{\ell 0}$ in ecliptic coordinates, not $Y_{\ell \ell}$
(see the zodiacal cloud images  in \cite{Dikarev2008}). Therefore, while the
zodiacal dust is unlikely to cause a lack of large angle correlation, it
could change the significance of some of the anomalies. 

Another Solar System source of CMB foreground might be
the Kuiper belt \cite{Hansen2012} and more Solar System
related ideas have been studied in \cite{Frisch2005}, proposing nearby
interplanetary dust towards the nose of the heliosphere to be responsible of
some of the unexpected alignments.

\subsection{Milky Way}

The next well-established layer of foregrounds are due to the Galaxy. At the highest frequencies
galactic thermal dust is dominant and molecular lines from CO transitions contribute in various 
frequency bands \cite{Planck2015DiffuseComponents}.

Until recently it was believed that at low frequencies synchrotron and free-free
emission are the dominant mechanisms. Interestingly enough, the Planck 2015
release overturned that point of view and showed that free-free and spinning
dust are the dominant components at the lowest frequencies \cite{Planck2015LowFrequencyForegrounds}.  

Shortly after the discovery of the low multipole alignments, one suspicion was
that they could be caused by residual contamination due to
Galactic foregrounds \cite{Slosar2004}. If the Galactic plane signal contributes 
to the large-angle CMB, multipole vectors should point in the direction of the plane 
and, more generally, alignments should be Galactic (and not largely ecliptic).  
This has been explicitly demonstrated by \cite{CHSS2006}, who found that adding a
Galactic-emission-shaped template contributing to the CMB map with an
arbitrary weight does not lead to observed alignments. The multipole
vector analysis is particularly effective in this case, as it
easily detects the directions singled out by the Galaxy, i.e. the Galactic
center and the Galactic poles.

It also seems to be hard to explain a lack of correlation at the largest
angular scales from residual galactic contamination. The galaxy is quite close
to us and thus highly correlated over large scales (it extends over much more
than $60$ degrees on the sky).  The comparison of the two-point correlation
function for different masks, as well as constraining the analysis to correlations for
which at least one point is close to the Galactic plane, are in full agreement
with the hypothesis that foreground cleaned full-sky maps do contain Galactic
residuals that strongly affect the amount of correlation at the largest
angular scales. However the fact that the most conservative masks provide the
smallest amounts of correlation seems to indicate that
it is precisely the cleanest, most trustworthy regions on the sky that
show the strongest evidence for the vanishing of angular correlations in the CMB
%% the lack of correlation cannot be explained by a Galactic foreground
\cite{CHSS2009,Gruppuso2014}.

%% \dragan{[Aren't the last two sentences mixing
%%     lack of correlations in the full sky, vs that on the cut sky? When we let
%%     one `leg' of the correlation be in the Galactic plane, we found nonzero
%%     correlation, but not when both legs are outside of the plane/mask.]}
%% \dominik{What I wanted to say is that if you include more sky you get more correlation, which is
%% an argument against the possible hypothesis that the lack of correlation could be due to the 
%% galaxy. It shows that the galaxy actually adds to the correlation. Maybe it needs rephrasing.}

Recently it became clear that there is another type of galactic foreground
that seems to be more local and might have a quite complex structure. The
so-called radio loops are believed to be relics of supernovae.  They have been
detected at radio frequencies long ago, and have been believed to be of no
relevance for the CMB temperature anisotropies. However, it was argued,
especially in the context of polarized emission, that this might not be true
\cite{MertschSarkar2013,Liuetal2014,Ogburn2014,Vidaletal2015,Planck2015DiffuseComponents}. Most
recently, it was shown that a loop structure in the vicinity of the cold spot,
together with another structure called
radio loop I, is able to almost perfectly reproduce the observed
quadrupole-octopole map \cite{PhDFrejsel}.  If indeed these two structures
would dominate the sky at the very low multipole moments, then the primordial
fluctuations at those scales must be completely absent. This might explain the
alignments and maybe to some extend the dipolar modulation, but the lack of
correlation would become more significant and would be in stark contrast to
the $\Lambda$CDM model.

\subsection{Other foregrounds} 

There are a number of other foregrounds that could in principle have
effect on, or even be the cause of, the anomalies. For example, the local
extragalactic environment in form of hot plasma and a local Sunyaev-Zeldovich
effect may play a role \cite{Abramo2006-anomalies,Peiris2010}. 
However, these other foregrounds have not been studied in great detail in the context of the 
anomalies, partly because they are not thought to be able to generate features at very
large angular scales.

The short summary of this section, therefore, is
that  while  none of the aforementioned effects have been proven to cause any of the
CMB anomalies, it is clear that these are physically well motivated
foregrounds and an improved understanding of them will also help us to better
understand the nature of the unexpected CMB features.

The major argument against a foreground related explanation of the CMB
anomalies is the frequency independence of the observed anomalies.  In fact,
the anomalies show up at more or less the same statistical significance in
four different Planck pipelines that lead to foreground cleaned maps of the
full sky and in the corresponding WMAP pipeline.  This implies that any so far
unidentified foreground that would be responsible for one or all of the
unexpected features of the CMB at large angular scales would have to mimic a
CMB fluctuation spectrum, in order not to show up in the difference maps
between the four foreground cleaned Planck maps (Commander, NILC, SEVEM,
SMICA).

\section{Cosmology}

Perhaps the most exciting possibility is that some or all of the anomalies have a
(common) cosmological origin. In this section, we consider a variety of proposed
cosmological mechanisms whose manifestation could be the observed anomalies.

\subsection{Kinetic effects}

%\dragan{[recommend global replace kinetic $\rightarrow$ kinematic in this section]}
% 

Earth's motion through the rest frame of the CMB leads to higher-order
effects on the observed anisotropy, which could in principle affect
conclusions about the observed anomalies \cite{Kamionkowski2003}. As already
discussed above, these so-called kinetic effects have been studied
for low multipole moments \cite{SSHC2004, CHSSalignments2015, Notari:2015kla})
as well as for the highest multipole moments \cite{PlanckProperMotion,
Planck2015Isotropy} and both contribute to the final significance for the anomalies.  
The kinetic effect on the quadrupole also provides another argument against a Solar System,
Galactic or local extragalactic foreground. When this well-understood
correction to the data is applied, evidence for the alignments becomes even stronger. If those
alignments were caused by, say, a Galactic foreground, the correct kinetic
correction should be derived from the velocity of the Solar System within the
Galaxy and not with respect to  the CMB frame. In that case a ``wrong'' kinetic correction 
would have been applied, which would be very unlikely to increase the alignments (a random 
correction actually leads most likely to a less significant alignment). This indicates that 
the alignment is a physical effect and that it is not due to foregrounds. 

\subsection{Local large scale structure} 
\label{sec:local_LSS}

Local structure -- over/underdensities in the dark matter distribution within
tens or few hundreds of megaparsecs of our location in the universe -- could
in principle be responsible for some of the alignments. This class of
explanation has a nice feature of producing large-scale effects relatively
easily, since the small distance to us implies a large angle on the sky (see
Fig.~\ref{AngularScales}).

One possibility is the late-time integrated Sachs-Wolfe effect (ISW), or, in the
non-linear regime the Rees-Sciama effect. This is the additional anisotropy
caused by the decay of gravitational potential when the universe becomes dark
energy-dominated (redshift $z\lesssim 1$), and has a nice feature that it is
achromatic.  First estimates showed that the effect could give rise to the
correct order of magnitude for the quadrupole and octopole
\cite{Rakic2006a,Inoue2006}.
%% , but \dragan{who? There is an old paper by Seljak and Slozar I think.}
%% \dragan{It's not them...} \dominik{I found the paper that I had in mind,
%%   Seljak, 1996 ApJ 460, 549, but thats just showing that the RS effect in the
%%   old CDM model is a $10^{-6}$ effect and peaks at $\ell \sim 100$. I think it
%%   is not in contradiction with the statements above. Maybe we should just skip
%%   the `but' part.} reached different conclusions.
In \cite{Rakic2006a} a single spherical structure was considered and it was
argued that a single over or underdensity cannot give rise to the observed
pattern. In \cite{Inoue2006} a more complicated configuration of voids was
considered and it was shown that such a structure could explain the observed
alignment.  This idea was studied further in
\cite{Francis2010,Dupe2011,Rassat_Starck_Dupe,Rassat2013}.

An argument against explaining the observed alignments with the ISW is simply
that it is unlikely: barring a suppression of primordial temperature
fluctuations, the observed missing power at large angles generically requires
a chance cancellation between the local ISW signal and the primordial CMB
pattern.  This is unlikely and, taken at face value, would imply another
anomaly \cite{CHSSalignments2015}.  Nevertheless, this idea can eventually be tested by means of
cross-correlation of the CMB maps with all-sky maps of the cosmic structure. 

The idea that an unusually large void is in our vicinity has been revived 
repeatedly in the context of the cold spot anomaly. There had
been a claim of an underdensity in the NVSS radio survey in the location of
the CMB cold spot \cite{Rudnick:2007kw}, but this feature was proven to not
be statistically significant once the systematic errors in the survey, and
in particular the known underdensity stripe in NVSS, are taken into account
\cite{Smith:2008tc}. More recently, there was a claimed discovery of a
large ($\sim 200$\,Mpc) underdensity ($\delta\rho/\rho\sim -0.15$) centered at
redshift $z\sim 0.2$ in the distribution of galaxies in the 2MASS-WISE survey
\cite{Szapudi:2014zha}. The underdensity lies in the direction of the CMB cold
spot, leading to a fascinating possibility that the former is causing the
latter via the ISW effect \cite{Kovacs:2014ooa}. However, this causal
explanation has been brought into question
\cite{Zibin:2014vaa,Nadathur:2014tfa}, as it appears that the underdensity is
not sufficiently pronounced to cause the observed temperature cold
spot. Future tests, discussed in Sec.~\ref{sec:way_forward}, will have a lot
more to say about the local structures and their relation to CMB anomalies.

\subsection{Primordial power spectrum --- broken scale invariance}

An inflationary scenario with the minimally short period of slow-roll,
say just 50 to 60 e-folds, could accommodate breaking of the scale
invariance during inflation at observationally accessible scales, that in
turn could manifest itself as one or more of the CMB anomalies
\cite{BdVS2006a, BdVS2006b, PowellKinney2007, NicholsonContaldi2008, DdVS2008,
RamirezSchwarz2009, RamirezSchwarz2012,
Ramirez2012,Cicolietal2014,Handleyetal2014,ScaccoAlbrecht2015,Gruppuso2015}.
Alternatively one could also consider scenarios in which inflation deviates
from its generic slow-roll behavior just 50 to 60 e-foldings before it ends,
e.g. \cite{Contaldietal2003,Jainetal2009}.  Many of those models find an
improved quality of fit to the observations, however the improvement is
typically not statistically significant given the additional
parameters of the model. These models also make definite predictions on
tensor modes and on the polarization of the CMB at the largest angular scales;
thus, there are more handles that we can hope to exploit in the
future. However, a generic problem of an explanation along these lines is that
a new fine tuning (why do we live in the epoch when the Universe is large
enough to observe the first pre-inflationary scales) is introduced in some of
these models.

\subsection{Primordial power spectrum --- broken isotropy}

Tests of isotropy and homogeneity of the initial conditions in the universe
have seen tremendous activity and development over the past decade. Two facts
contributed to this. First, several of the anomalies, particularly the parity
anomaly and the hemispherical anomaly, can be naturally modeled (and
potentially explained) using modulation of either the primordial temperature
or primordial power spectrum. Second, the advent of full-sky WMAP and Planck
CMB maps enables the precise measurements required to constrain these models,
particularly the high-$\ell$ couplings between the  multipoles.

At the level of the CMB temperature, the modulation can most generally be
written as \cite{Gordon2005}
\begin{equation}
  T(\unitvec{e}) = T_0(\unitvec{e})[1+ f(\unitvec{e})],
  \label{eq:dipole_genmod}
\end{equation}
where $f$ is some function. Ref.~\cite{Gordon2005} studied both the dipolar and
the quadrupolar
temperature modulation (i.e.\ when $f$ is proportional to $Y_{1m}$ and $Y_{2m}$,
respectively) in order to explain the missing correlations at large
scales and the quadrupole-octopole alignment. While it is certainly possible
to do so, \cite{Gordon2005}  found that it is difficult to {\it naturally} arrange
for the missing correlations, as it typically requires chance cancellations at
large scales.

The discovery of hemispherical asymmetry gave much further impetus to the
study of modulations, particularly the dipolar one written down in
Eq.~(\ref{eq:dipole_mod}). It has become particularly interesting to ask what
general mechanisms could produce a such a long-wavelength modulation. An
inflationary theory could, in principle, accommodate models that produce
hemispherical asymmetry, but such a model would have to be multi-field and
involve, for example, a large-amplitude superhorizon perturbation to the
curvaton field \cite{Erickcek:2008sm}.  It turns out that a long wavelength,
superhorizon-scale gradient in density (the so called Grischuk-Zeldovich
effect \cite{grishchuk1978long}) could not produce the dipolar asymmetry
starting with adiabatic fluctuations because the intrinsic dipole in the CMB
produced by the perturbation is exactly canceled by the Doppler dipole
induced by our peculiar motion
\cite{turner1991tilted,Bruni:1993dx,Erickcek:2008jp}. However, a superhorizon
{\it isocurvature} perturbation could do the job
\cite{Erickcek:2008jp}. In that case, the effects of the superhorizon
fluctuation would also presumably be seen as the anisotropic distribution
of large-scale structure on the sky, but no such effect has yet been
detected in the distribution of quasars \cite{Hirata:2009ar} or galaxies
and other tracers \cite{Pullen:2010zy,GibelyouHuterer2012}. 

An equally interesting possibility is that anisotropic inflation led to a
breaking of statistical isotropy. The best-studied model posits that the power
spectrum take the following form \cite{Ackerman:2007nb}
\begin{equation}
  P(\vec{k}) = P(k) [1 + g_* (\vec{k}\cdot \unitvec{d})^2]
  \label{eq:ACW}
\end{equation}
where $\unitvec{d}$ is again a special direction. Such a model would imply
that inflation was anisotropic, which could be caused by coupling to vector
fields \cite{Soda:2012zm}, presence of magnetic fields
\cite{Kahniashvili:2008sh} or models motivated in supergravity
\cite{Kanno:2010nr}. First estimates of the parameter $g_*$ found it is
nonzero at the huge significance of $\sim 9\sigma$ \cite{Groeneboom:2008fz},
but this was soon found to be due to a known effect of asymmetric beams
\cite{Hanson:2010gu} which had not been taken into account. A later analysis
based on Planck data gave the best constraint to date, $g_*=0.002\pm 0.016$ at
68\% C.L \cite{Kim:2013gka}; the Planck team gets very similar 
constraints~\cite{Planck2015Isotropy,Planck2015Inflation}.

It is also interesting that the apparent breaking of statistical isotropy can
actually be an artifact of non-Gaussianity
\cite{Byrnes:2011ri,Schmidt:2012ky}. More precisely, coupling between a
superhorizon long mode and shorter, observable modes due to primordial
non-Gaussianity on superhorizon scales can manifest itself in observations as
a preferred direction on the sky. Essentially, one can thus ``trade'' the
breaking of statistical isotropy for the presence of primordial
non-Gaussianity. These ideas have been further elaborated by
\cite{Nelson:2012sb,LoVerde:2013xka,Adhikari:2015yya}.

\subsection{Topology} 

A non-trivial topology of the Universe might in principal both lead to a lack
of correlation at large angular scale and introduce alignments and/or
asymmetries, while preserving a locally isotropic and homogeneous geometry.
The idea is that the Universe might have a finite size which is not much
larger than today's Hubble distance.  In such a universe there is a natural
cut-off for structures at large scales \cite{Aurich2004,Aurich2005,Aurich2008}
and if the different large but compact dimensions are not of equal size, we
could even imagine that a plane like the quadrupole-octopole plane would be
singled out \cite{Aurich2007,BR2009}.

Detailed studies of non-trivial topologies however did not find any
statistically significant signal to substantiate those ideas
\cite{Cornish:2003db,ShapiroKey:2006hm,Vaudrevange:2012da,Planck2015GeometryTopology}.
These included generic studies based on the circles-in-the-sky signature
\cite{Cornish:1997ab}, which rule out with reasonable confidence that there is
a non-trivial closed loop in the universe with length less than 98.5\% of the
diameter of the last scattering surface \cite{Vaudrevange:2012da}.  Searches
for anomalous correlations in the CMB can extend these bounds for specific
manifolds or classes of manifolds to slightly larger distances.

For non-trivial topology to be observable, whether or not it is behind any of
the mysterious large-scale features in the CMB, the characteristic length
scale of the fundamental domain would need to be comparable to the Hubble
distance.  We would then be faced with another coincidence problem -- why do
we live in an epoch in which we are able to see a non-trivial topology?

A short summary of the cosmological explanations offered so far is that the
ideas mentioned here could all explain at least one ``anomaly atom'', however,
none of them has been demonstrated to be detectable at a statistically
significant level. The kinetic effects must be taken into account, but do not
seem to explain any of the new features of the CMB.  The local large scale
structure exists and must be better understood -- especially form non-CMB
observations (see below). Primordial physics might lead to broken scale
invariance, very large non-adiabatic modes and non-trivial topologies. All
three suffer from a coincidence problem.  This problem might be most prominent
for finite topologies, like a 3-torus. When combined with inflation such a
solution requires a single short period of inflation. All ideas of primordial
power suppression suffer from a regeneration of large scale power via the ISW
effect at late times. This aspect is tamed in finite topologies, but in that
case the alignment of modes is diluted via the ISW effect
\cite{BR2009}. Finally the idea to break the isotropy of the
primordial power spectrum seems -- when compared to data -- not to be
implemented in the CMB on all scales (in particular there is no evidence for
such an effect at $\ell \gsim 60$) and thus must be combined with a breaking
of scale invariance. On top it is unclear how such a primordial breaking of
isotropy would generate a lack of correlation at large angular scales.

\section{Way forward and Conclusion}
\label{sec:way_forward}

The origin and nature of CMB temperature anomalies can be tested with
other observations in cosmology. The best way to proceed along those lines is to assume the
model for the CMB temperature anomalies, and test that model with other data. 

The simplest model that can be tested is that the anomalies are simply fluke
events in the standard $\Lambda$CDM cosmological model. In the language of frequentist
statistics, the anomalies in this case are realized in a very small fraction of random
realizations within the underlying $\Lambda$CDM model. Then one could simply ask how
the other observations, beyond CMB temperature, are affected as seen in
this small subsection of anomalous $\Lambda$CDM realizations. One could similarly
test other, non-$\Lambda$CDM models for the anomalies whether they are fundamental or
purely phenomenological and, if desired, use Bayesian statistics as well. 

We now discuss how other observations in cosmology can help understand the CMB
temperature anomalies.

\subsection{CMB polarization}

Any given model for the CMB temperature anomalies has predictions for
polarization. Given the increasingly precise polarization measurements, it is
of great interest to make predictions for polarization under different models
for the anomalies. The polarization data from WMAP and Planck are good enough
to produce reliable polarization power spectra, but are not high enough
signal-to-noise to produce {\it maps} of the polarization field -- especially
not at very large scales where foregrounds are extremely difficult to
remove. Nevertheless, new generations of experiments, such as LiteBird
\cite{Matsumura:2013aja}, CORE \cite{Bouchet:2011ck}, and CMB-S4 hold promise
for accurate maps of polarization down to the lowest
multipoles. 

First predictions for polarization were carried out by \cite{Dvorkin:2007jp}
who studied how a dipolar modulation model that can explain the hemispherical
asymmetry and a quadrupolar modulation model that can explain the
quadrupole-octopole alignment can be constrained with polarization map data.
For the dipolar case, they showed that predictions for the correlation between
the first 10 multipoles of the temperature and polarization fields can
typically be tested at better than the 98\% CL; while for the quadrupolar
case, predicted correlations between temperature and polarization multipoles
out to $\ell = 5$ provide tests at the 99\% CL or stronger.  This was followed
up by \cite{Copi:2013zja} who assumed that the suppressed correlations
observed in WMAP and Planck temperature data are just unlikely realizations in
the $\Lambda$CDM model, and studied what that hypothesis predicted for CMB
temperature-polarization cross-correlation.  Their conclusion is that while
the temperature-polarization cross-correlation cannot definitively be expected
to be able to rule out $\Lambda$CDM, one can nevertheless construct statistics
that have a good chance ($\sim 50\%$) of excluding the hypothesis at a high
statistical confidence ($>3\sigma$).  Similar results were obtained
\cite{Yoho:2013tta} for the predicted utility of large-angle
temperature-lensing-potential correlation function measurements; however, in
\cite{Yoho:2015bla} the authors showed that the large-angle polarization
auto-correlation function may be more promising.  The latter also began
exploring the alternative hypothesis that the lack of large-angle temperature
auto-correlation reflects a lack of large-distance correlations in the metric
potentials, and note that the large angle polarization auto-correlation
appears well suited to test this hypothesis.

\subsection{Large-scale structure}
\label{sec:LSS}

The distribution of galaxies, or other tracers of the large-scale structure,
may be particularly useful to clarify the nature of at least some of the
observed CMB anomalies. Galaxies have now been mapped over the whole sky out
to $z\sim 0.2$, e.g.\ by the 2MASS \cite{Skrutskie:2006wh} and Wide Infrared
Survey Explorer (WISE; \cite{Wright:2010qw}) surveys, and to a depth of $z\sim
0.7$ by the SDSS, including a subsample of more than two million galaxies and
quasars that have spectroscopic information \cite{Alam:2015mbd}. In the
future, a combination of the Dark Energy Survey (DES; \cite{Abbott:2005bi}),
Dark Energy Spectroscopic Instrument (DESI; \cite{Levi:2013gra}), Euclid
\cite{Laureijs:2011gra}, Large Synoptic Survey Telescope (LSST;
\cite{Abell:2009aa}) and Wide-Field Infrared Survey Telescope (WFIRST;
\cite{Spergel:2015sza}) will map out the galaxy distribution over the whole
sky out to redshift beyond one, while eROSITA \cite{Merloni:2012uf} will carry
out a complete census of X-ray clusters in the observable Universe.  Other
tracers, such as quasars and radio galaxies, are particularly useful since
they are at redshifts of a few and probe an even larger volume. Eventually HI
surveys, e.g. by means of the Square Kilometre Array (SKA;
\cite{Schwarz:2015pqa}), will map all galaxies containing neutral hydrogen out
to a redshift of a few. The SKA will also allow us to eventually pin down the cosmic 
radio dipole at per cent level and thus provide a precise test of 
the proper motion hypothesis of the CMB dipole and possibly identify a significant structure 
dipole \cite{Schwarz:2015pqa}. 

To give another example, missing power on large angular scales leads, assuming $\Lambda$CDM, to
corresponding suppression of power on a gigaparsec scale (that is, the power
spectrum of matter fluctuations $P(k)$ is suppressed for $k\lesssim
0.001\hmpcinv$). It turns out that a future very-large volume survey, such as
LSST, could in principle measure power on such large scales to a sufficient
accuracy to detect the purported suppression in power at a statistically
significant level \cite{Gibelyou:2010qe,Hearin:2011xv}. 

The cold spot offers a particularly appealing target for the large-scale
structure surveys because of its smaller angular size. While the
aforementioned evidence for the underdensity in the large-scale structure is
inconclusive (see Sec.~\ref{sec:local_LSS}), the aforementioned future wide,
deep surveys will offer a fantastic opportunity to correlate the large-scale
structure to the CMB anomalies. 

\subsection{Foregrounds}

From the discussion in Sec.~\ref{foregrounds}, it is clear that both
Solar System and Galactic foregrounds need to be better understood.
For example, we already mentioned that three different models for
the zodiacal dust emission mutually disagree by more than an order of
magnitude.  It would be important to investigate the prospect of
dedicated observations targeted specifically in regions
% closer to the Sun,
where the zodiacal dust signal becomes more important to test those
models. Similarly, the nature of spinning dust is not well
understood (see e.g.~\cite{Draine2012,Hensley2015}). The same holds for the extrapolation of known radio loops into the
relevant microwave bands. In both cases, a dedicated radio and microwave
observations would be useful. The DeepSpace project in Greenland is planning
to cover those frequency bands. More generally, other astrophysical observations
in the coming 10--20 years should be able to provide significant new information
about foregrounds and their effects on the CMB anomalies.

\section{Executive summary}

We summarized the evidence that several CMB anomalies are real
features. We identify three anomaly ``atoms'': the lack of large angle
correlation, the mutual alignment of the lowest multipole moments, and the
hemispherical asymmetry. These three ``atoms'' seem to be orthogonal and
independent of each other in the realm of the minimal inflationary
$\Lambda$CDM model. Any proposal to explain (or just parameterize) these new
CMB features should address at least two of the ``atoms'' as one of
them might, after all, still be a statistical fluctuation. In an
effort to find a compelling explanation, several new theoretical ideas have been
considered, covering a broad spectrum from the physics of dust grains to
exotic theories of the very early universe.  Currently, the physics
behind the CMB anomalies is still unknown, but new observations of the 
CMB (especially of polarization) and new observations of at other wavebands, 
both of the large scale structure and of potential foregrounds, will provide significant
new information and provide us with powerful new tools to eventually resolve the
puzzle of the anomalies.

\ack

DJS is supported by the DFG grant RTG 1620 `Models of Gravity'.  GDS and CJC
are supported by the US Department of Energy grant DOE-SC0009946.  DH is
supported by NSF under contract AST-0807564 and DOE under contract
DE-FG02-95ER40899.

We acknowledge the use of the
Legacy Archive for Microwave Background Data Analysis (LAMBDA), part of the
High Energy Astrophysics Science Archive Center (HEASARC). HEASARC/LAMBDA
is a service of the Astrophysics Science Division at the NASA Goddard Space
Flight Center.  This paper made use of observations obtained with
Planck\ (www.esa.int/Planck), an ESA science mission with
instruments and contributions directly funded by ESA Member States, NASA,
and Canada.

\section*{References}

\bibliographystyle{iopart-num}
\bibliography{CQGAnomalies}

\end{document}